\documentclass[a4paper,11pt]{article}

\usepackage[utf8]{inputenc}
\usepackage[T1]{fontenc}

\usepackage{amsmath,amssymb,amsfonts}
\usepackage{graphicx}
\usepackage{mathtools}
\usepackage{bm}
\usepackage{booktabs}
\usepackage{dcolumn}
\usepackage{braket}
\usepackage{babel}
\usepackage{cite}
\usepackage{float}
\usepackage{geometry}
\usepackage{enumitem}
\usepackage{xcolor}
 \usepackage{hyperref}
\usepackage{cleveref}

\begin{document}
\begin{center}

{\Large{\bf Influence of Perfect Fluid Dark Matter on Shadow Observables of Yang-Mills modified charged black holes}}

\vspace{8mm}

\renewcommand\thefootnote{\mbox{$\fnsymbol{footnote}$}}
Abhishek Negi  ${}^{1}$\footnote{spacetime122002@gmail.com}
Md Sabir Ali,${}^{2}$\footnote{alimd.sabir3@gmail.com}
Shagun Kaushal,${}^{3}$\footnote{shagun.kaushal@vit.ac.in}
Sanjay Pant,${}^{4}$\footnote{sanjay.pant@dituniversity.edu.in (corrresponding author)}

\vspace{4mm}

 ${}^1${\small \sl Department of Allied Sciences (Physics), \sl Graphic Era (Deemed to be University), \sl  Dehradun, Uttarakhand 248002, India} 
\vskip0.2cm
${}^2${\small \sl Department of Physics, \sl Mahishadal Raj College, \sl  West Bengal 721628, India} 
\vskip 0.2cm
${}^3${\small \sl Department of Physics, \sl School of Advanced Sciences, \sl Vellore Institute of Technology,
Vellore 632014, India} 
\vskip 0.2cm
${}^4${\small \sl Department of Physics, School of Physical Sciences, \sl DIT University, \sl Mussoorie-Diversion Road, Makkawala, Dehradun, Uttarakhand -248009, India} 
\end{center}

\vspace{6mm}
\numberwithin{equation}{section}
\setcounter{footnote}{0}
\renewcommand\thefootnote{\mbox{\arabic{footnote}}}
\begin{abstract}
We investigate the influence of perfect fluid dark matter (PFDM) on Yang--Mills--inspired charged black holes, with a particular focus on the resulting modifications to key black hole observables. By embedding a PFDM term into the spacetime geometry, we examine the alterations in shadow morphology, photon geodesics, and the associated energy emission spectra. Our analysis reveals that PFDM induces notable deviations in the shadow size, shape, and circularity, and significantly impacts the stability of circular orbits. Furthermore, the energy emission rate exhibits a strong dependence on both the Yang--Mills charge and the dark matter distribution. These results indicate that environmental effects arising from dark matter can imprint observable signatures on black hole shadows and radiation processes, offering a potential pathway to constrain dark matter models and probe non-Kerr geometries with forthcoming high-precision observations such as those from the Event Horizon Telescope and next-generation interferometers.
\end{abstract}

\newpage
\tableofcontents
\newpage

\section{Introduction}
The recent surge of interest in black hole physics has been propelled by groundbreaking observational achievements such as the imaging of the supermassive black hole M87* by the Event Horizon Telescope (EHT) collaboration and the detection of gravitational waves from black hole mergers by LIGO, Virgo, and KAGRA \cite{EventHorizonTelescope:2019dse,LIGOScientific:2016aoc,LIGOScientific:2018mvr,KAGRA:2021vkt}. These observations not only confirm key predictions of general relativity but also open an invaluable window into the strong-field regime of gravity, where potential deviations from the classical theory and environmental effects could manifest. As a result, the need to develop and analyze alternative black hole models, including those arising from extended theories of gravity or modified matter sectors, has become increasingly pressing.

Among the various avenues of exploration in gravitational physics, black holes coupled to non-Abelian gauge fields—particularly those sourced by Yang–Mills fields—constitute a remarkably rich and intriguing class of solutions, often exhibiting properties distinct from their Abelian counterparts \cite{Bizon:1990sr,Volkov:1998cc,Winstanley:2008ac}. Unlike the Abelian electromagnetic field, non-Abelian Yang–Mills fields exhibit intrinsic nonlinearity due to their self-interacting gauge structure. This nonlinearity leads to black hole solutions with qualitatively distinct causal structures, thermodynamic behavior, and stability properties. In particular, charged black holes inspired by Yang–Mills theory offer a rich intersection between classical general relativity and high-energy gauge theories, yielding potentially observable deviations from standard solutions like the Reissner–Nordström black hole \cite{Bartnik:1988am,Volkov:1998cc}.

Astrophysical black holes are inherently embedded within complex astrophysical environments, dynamically interacting with surrounding matter and radiation fields; hence, they cannot be accurately modeled as isolated systems \cite{Narayan:2008bv,Abramowicz:2011xu, Bambi:2017khi}. In the cosmological context, it is well established that most of the matter content of the universe is composed of dark matter, whose precise nature remains elusive \cite{Ishak:2025cay}. It is therefore natural to expect that the gravitational field of a realistic black hole would be influenced by the presence of a surrounding dark-matter halo. While various models attempt to capture the interaction of black holes with dark matter, one particularly tractable and phenomenologically appealing approach is the use of Perfect Fluid Dark Matter (PFDM). In this model, dark matter is modeled as an anisotropic perfect fluid, allowing for a consistent integration into black hole spacetime without resorting to specific particle microphysics \cite{Li:2012zx}.

Motivated by these developments, the present work aims to construct and analyze a new class of black hole solutions obtained by incorporating a PFDM component into the Yang-Mills-inspired charged black hole spacetime. This novel configuration enables a detailed examination of how the presence of dark matter alters key black hole observables, potentially leading to observable signatures in current and future astrophysical data.

To this end, we systematically investigate several critical aspects of the modified black hole geometry. In \cref{DMPFmdl} we obtain the charged black hole solutions with PFDM background in Yang-Mills theory. Next, we explore the geodesics of the light rays and subsequently analyze the shadow cast by the black hole, as an outcome of the unstable photon orbits, providing important insights into the black hole spacetime geometry. The deformation and size of the shadow can serve as potential indicators of the surrounding dark matter distribution, if any, and its possible interaction with gauge-charged black holes. In addition, we examine the geodesic structure of the spacetime, particularly focusing on circular orbits and their deviation under perturbations. Geodesic motion underpins gravitational lensing phenomena, accretion disk dynamics, and stellar motion near black holes, and hence its study in a PFDM-augmented Yang-Mills charged background is of significant astrophysical relevance. Moreover, we analyze the energy emission rate, crucial for understanding the Hawking radiation and thermodynamic properties of black holes in a more realistic setting influenced by ambient dark matter. 

The broader implications of this work are twofold. From a theoretical standpoint, it enriches the family of known exact solutions involving non-Abelian fields and cosmologically motivated matter components, offering new laboratories for probing the strong field regime of gravity. From an observational perspective, it provides modified signatures--in the form of altered black hole shadows, and changes in energy emission — that could potentially be used to infer the properties of the dark matter distribution around black holes. This is particularly timely given the forthcoming era of next generation observational facilities, such as the ngEHT campaigns, the Einstein Telescope, and LISA, which will greatly enhance our ability to probe the near horizon region of black holes and the structure of spacetime itself.
\cite{Abdujabbarov_2012,Afrin_2021,Ali_2025,ali2019shadowrotatingchargedblack,Atamurotov_2021,Ayon-Beato:1998hmi,Boillat:1970gw,Crisnejo_2019,Iyer:1986np,jha2023superradiancestabilityrotatingcharged,Jusufi_2020,Jusufi_2020n, kiselev2003quintessentialsolutiondarkmatter,Kleihaus_2002,kokkotas1988black,Konoplya_2003,Konoplya:2003ii,Langlois:2021aji,Li_2012,Ma_2024,Nomura:2021efi,PhysRevD.81.124045,Rahaman_2010,sánchez2024shadowrenormalizationgroupimproved,schutz1985black,tu2025yangmillsfieldmodifiedrn,vázquez_esteban_2004,Wei_2013,Zhang_2006,Zhidenko_2003,Johannsen_2012,Kala_2025,Hod_2003, Konoplya:2019sns,Konoplya:2025mvj}

 Motivated by these observational prospects, the present analysis is also intended to examine whether the combined effects of the Yang-Mills charge and the PFDM parameter lead to observational signatures that are distinguishable from those of the Kerr spacetime. In particular, we investigate how these parameters modify the photon sphere, shadow morphology, and shadow observables that may be tested using current and future high-resolution black hole imaging observations.

The paper is organized as follows: In \cref{DMPFmdl}, we briefly review the Yang-Mills inspired charged black hole solution and introduce its extension that incorporates PFDM. In \cref{GeoCal} we present the study of geodesics, circular orbits, and the associated deviation parameters. In \cref{shadow} the analysis of the black hole shadow, the impact of the parameters that come from introducing the PFDM term has also been incorporated.
 In \cref{ObBHpara}, we compute the observables and put constraints on the parameters characterizing the black hole, estimating their energy emission rate and relevant astrophysical implications. In \cref{sag} we model our black holes with the observational constraints coming from of $\text{M87}$ and $\text{SgrA}^*$ black holes. In \cref{eer} we study the energy emission rate and show the effect of Hawking radiation on the energy spectra.  Finally, in \cref{summary} we summarize our results and conclude the paper.

\section{Black hole spacetime with perfect fluid dark matter }\label{DMPFmdl}
The Yang-Mills inspired charged black hole solution was discussed earlier in the literature \cite{Tu:2025zeb}. In addition, we add an extra term comprising  of PFDM environment. Such solutions affect the geometric as well as the physical properties of a black hole. Hence, for the black hole solution of our interest, we consider the action comprising the electromagnetic field, the Yang-Mills field, and PFDM minimally coupled to gravity. 

We start with the action as follows
\begin{equation}
    \mathcal{S} = \frac{1}{2} \int d^4x   \sqrt{-g}\left(R - \mathcal{L}_{\text{EM}} -\mathcal{L}_{\text{YM}}-\mathcal{L}_{\text{PFDM}}\right),
\end{equation}
where, 
$\mathcal{L}_{\text{EM}}$ = $F_{\mu\nu}F^{\mu\nu}$, $\mathcal{L}_{\text{YM}}$ = $F^{\text{YM}}_p=\text{Tr}\left(\mathcal{F}^{(a)}_{\alpha\beta}\mathcal{F}^{(a)\alpha\beta}\right)^p$ are respectively, the Maxwell and Yang-Mills field invariants, and $\mathcal{L}_{\text{PFDM}}$ being the Lagrangian of the PFDM.\\
Varying the action with respect to the metric tensor $g_{\mu\nu}$, we have the modified Einstein's field equations as
\begin{equation}
G_{\mu\nu}=R_{\mu\nu}- \frac{1}{2}R g_{\mu\nu} = \mathcal{T}^{EM}_{\mu\nu} + \mathcal{T}^{YM}_{\mu\nu} + \mathcal{T}^{PFDM}_{\mu\nu}
\label{EOMS}
\end{equation}
where,
$$\mathcal{T}^{\text{EM}}_{\mu\nu} =2\left(\mathcal{F}_{\mu\alpha}\mathcal{F}^{\alpha}_{\nu}  - \frac{1}{4}g_{\mu\nu} \mathcal{F_{\alpha\beta}}\mathcal{F^{\alpha\beta}}\right)$$
$$\mathcal{T}^{YM}_{\mu\nu} =2\left(\mathcal{F}^{(a)}_{\mu\alpha}\mathcal{F}^{(a)\alpha}_{\nu}-\frac{1}{4}g_{\mu\nu} \mathcal{F}^{(a)}_{\alpha\beta}\mathcal{F}^{(a)\alpha\beta}\right)$$

$$\mathcal{T}^{(\text{PFDM})\nu}_{\mu}=\text{diag}\left(-\rho_{\text{PFDM}},P_r,P_\theta,P_\phi\right),$$
where the energy density and pressure components satisfy $\rho_{\text{PFDM}} = -P_r$ and $P_\theta = P_\phi = \rho_{\text{PFDM}}/2$, with
$$-\rho_{\text{PFDM}}=\frac{\alpha}{8\pi r^3},$$ 
and $\alpha$ is the PFDM parameter \cite{Ma:2024oqe}.

To solve the field equations, Eqs.~(\ref{EOMS}) in a spherically symmetric static spacetime, we consider the four-dimensional metric $ansatz$, viz.,

\begin{equation}
\label{metrids}
ds^2=-f(r) dt^2+\frac{dr^2}{f(r)}+r^2(d\theta^2+\sin^2\theta d\phi^2),
\end{equation}
After solving the $(t,t)$-component of Eq.~(\ref{EOMS}) with the energy-momentum tensors, we obtain the metric function in the following form.

\begin{equation} \label{metricf} 
f(r)=1-\frac{2M}{r}+\frac{Q^2}{r^2}+Q_{YM}r^{2-4p}+\frac{\alpha}{r} \ln\left(\frac{r}{|\alpha|}\right). 
\end{equation}

Next, we discuss the rotating counterpart of the black hole solution, Eq.~(\ref{metrids}) comprising the metric function in Eq.~\ref{metricf}. The usual method to generate the rotating spacetime is constructed using the Newman-Janis algorithm \cite{Newman:1965tw,Newman:1965my}. A more robust and useful technique for generating rotating spacetime was developed in \cite{Azreg-Ainou:2014pra} by partly employing the complexification methods. We start with the seed spacetime metric (\ref{metrids}) and applying Azreg-A\"inou's modified Newman-Janis algorithm, and get rotating black hole solutions. The algorithms to generate the rotating spacetime from a generic static seed metric are discussed in details in \cref{appendix}. This metric is characterized by mass ($M$), spin parameter $a$, the Maxwell charge ($Q$) and Yang-Mills parameter ($Q_{\text{YM}}$) as well as surrounding PFDM parameter ($\alpha$).\\

The spacetime metric for rotating Yang-Mills modified charged black holes surrounded by PFDM in the usual Boyer–Lindquist coordinates $(t,r,\theta,\phi)$ reads as
\begin{equation}
\label{rotmetric}
ds^2 = -\frac{\Delta}{\rho^2}\left(dt-a\sin^2\theta d\phi\right)^2 +\frac{\rho^2}{\Delta}dr^2 +\rho^2d\theta^2+\frac{\sin^2\theta}{\rho^2}\left(adt-\left(r^2+a^2\right)d\phi\right)^2
\end{equation}
where,
\begin{equation}
\label{grr}
\Delta=r^2-2Mr+Q^2+Q_{YM} r^{4-4p}+a^2+\alpha\; r \ln\left(\frac{r}{|\alpha|}\right)
\end{equation}

\begin{figure}[H]
    \centering
    \includegraphics[width=0.45\linewidth]{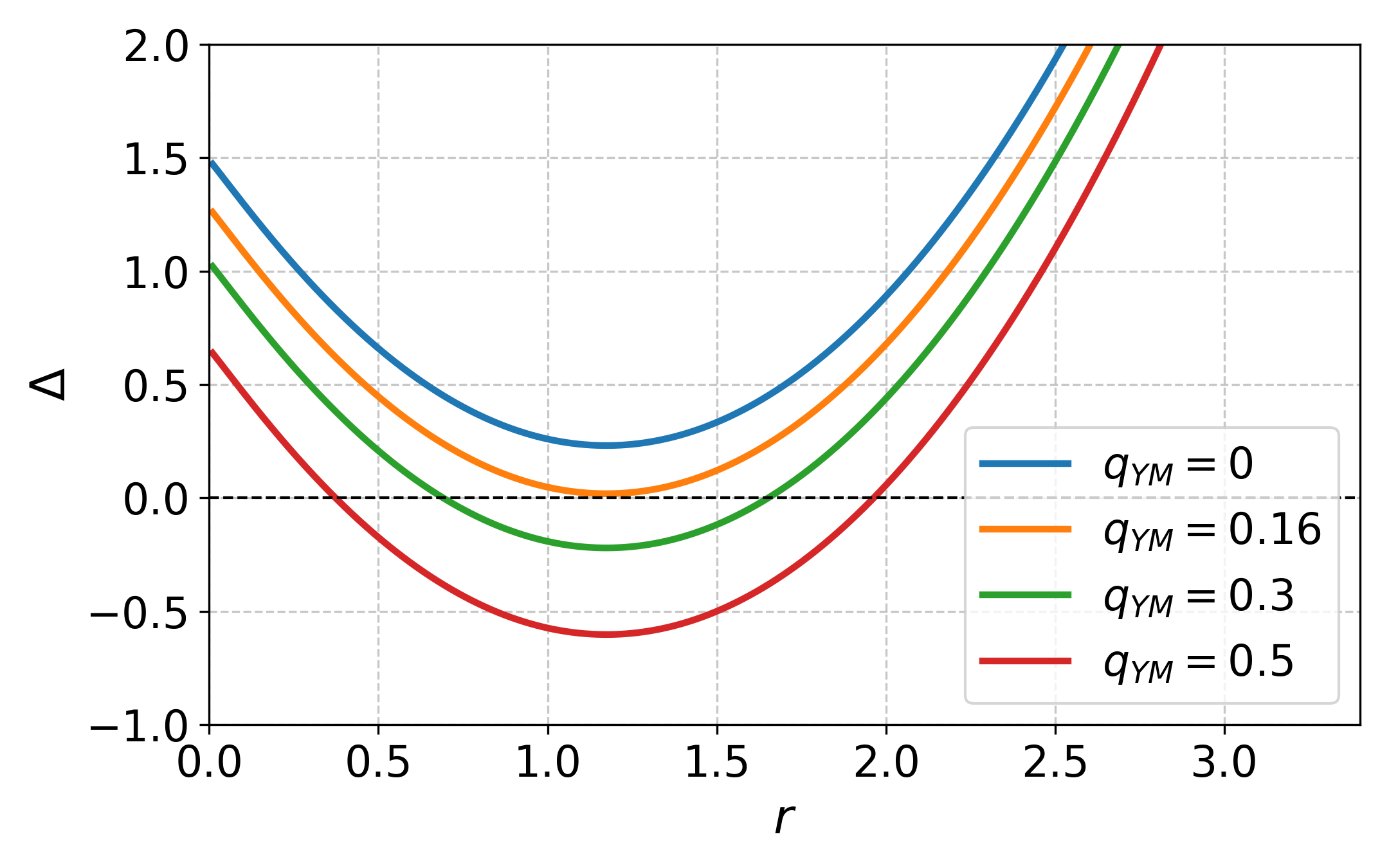}
    \includegraphics[width=0.45\linewidth]{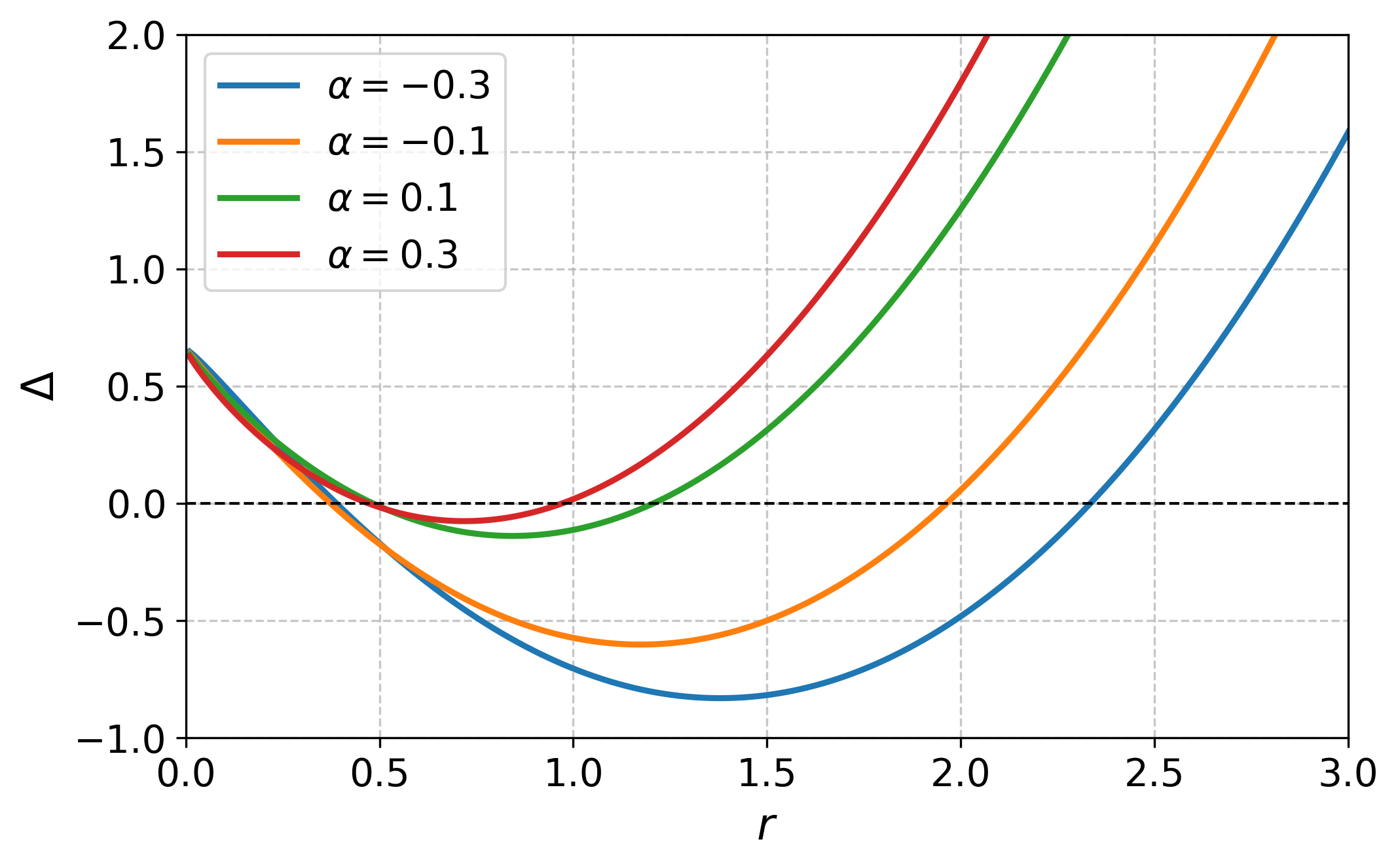}
    \caption{  The behavior of the $g^{rr}=\Delta$ vs the radial coordinate (r) for a set of values of $q_{\text{YM}}$ (\textit{\textbf{Left}}) and a set of values of $\alpha$ (\textit{\textbf{Right}}). Each of the plot admits two distinct horizons, namely, the inner horizon (Cauchy horizon) and the outer horizon (black hole event horizon). }
    \label{fig:horizon}
\end{figure}

\noindent The metric in Eq.~(\ref{rotmetric}) is an axially symmetric stationary spacetime and hence does not depend on time ($t$) and azimuthal $(\phi$) coordinates. This leads to having two Killing vectors $\xi^{\mu}_{(t)}=\delta^{\mu}{_t}$ and $\xi^{\mu}_{(\phi)}=\delta^{\mu}{_\phi}$, corresponding to the time translation and rotational invariance, respectively. These two vectors satisfy the Killing equation, $\xi_{\mu;\nu}+\xi_{\nu;\mu}=0$. The various metric components of the spacetime metric (\ref{rotmetric}) are evaluated as the scalar products of these killing vectors.
\begin{eqnarray}
\label{Nmcomp}
\xi^{\mu}{_{(t)}}\xi_{(t){_\mu}}&=&g_{tt}=1 -\frac{2Mr}{\rho^2}+\frac{Q^2+Q_{YM}r^{4-4p}}{\rho^2} +\frac{\alpha}{\rho^2} {\ln\left(\frac{r}{|\alpha|}\right)},\\
\xi^{\mu}{_{(t)}}\xi_{(\phi){_\mu}}&=&g_{t\phi}=\frac{a\sin^2\theta}{\rho^2}\left(2Mr-Q^2-Q_{\text{YM}}r^{4-4p}-\alpha r\ln\left(\frac{r}{|\alpha|}\right)\right),\\
\xi^{\mu}{_{(\phi)}}\xi_{(\phi){_\mu}}&=&g_{\phi\phi}=\frac{\sin^2\theta}{\rho^2}\left((r^2+a^2)^2-\Delta a^2\sin^2\theta\right).
\end{eqnarray}
Next, we investigate the horizon structure of the rotating black hole spacetime (\ref{rotmetric}) by computing the value of the radial coordinate using the relation $g^{rr}=\Delta=0$. An analytical solution is intractable; however, numerical analysis demonstrates the existence of two distinct real roots. For our convenience, we take $M=Q=1$ and vary the Yang-Mills charge $q_\text{YM}$ and the PFDM parameter $\alpha$. From now on, we set the scale-free parameters $q_{\text{YM}}/M\to q_{\text{YM}}$, $\alpha/M\to \alpha$, $Q/M\to Q$, and $a/M\to a$. We depict in Fig.~\ref{fig:horizon} the horizon structure of the rotating spacetime for the Yang-Mills field-inspired charged PFDM black hole for various values of $q$ and $\alpha$. We plot in Fig.~\ref{fig:horizon} (left) the horizon structure for different values at a fixed value of $\alpha$. We observe that there exist three different cases--(i) two distinct horizons, namely the Cauchy (inner) and black hole event (outer) horizons when $q_{\text{YM}}\geq 0.3$; (ii) a degenerate horizon at $q=0.16$ when inner and outer horizons coincide; and (iii) a naked singularity when $q_{\text{YM}}<0.16$. A similar structure is obtained on the right plot of Fig.~\ref{fig:horizon}.

\begin{figure}[H]
    \centering
    \includegraphics[width=0.45\linewidth]{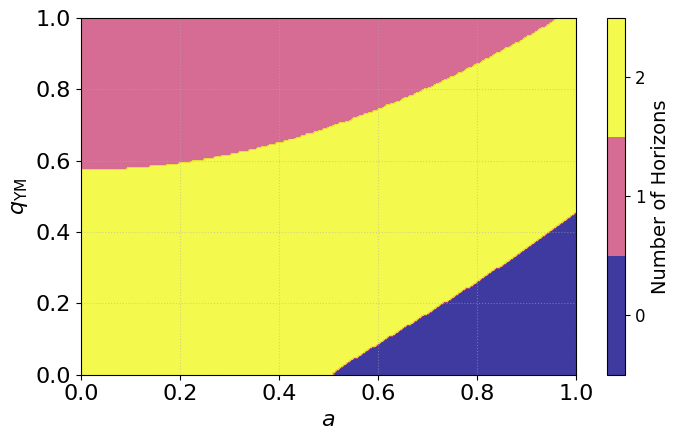}
    \includegraphics[width=0.45\linewidth]{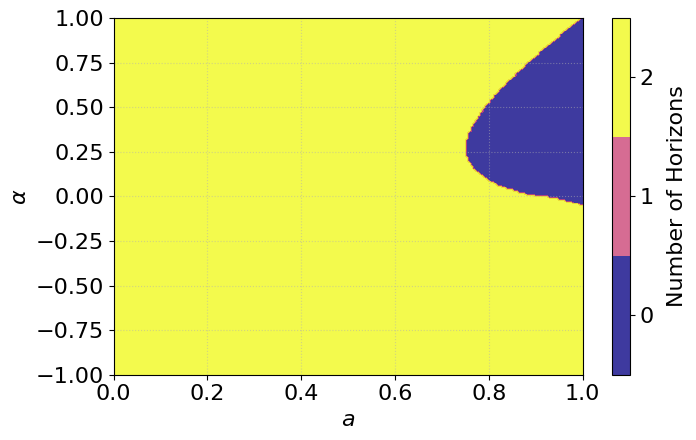} \\
    \includegraphics[width=0.45\linewidth]{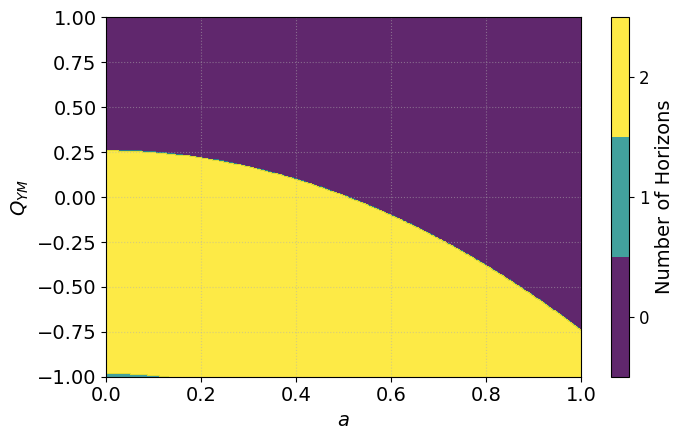}
    \includegraphics[width=0.45\linewidth]{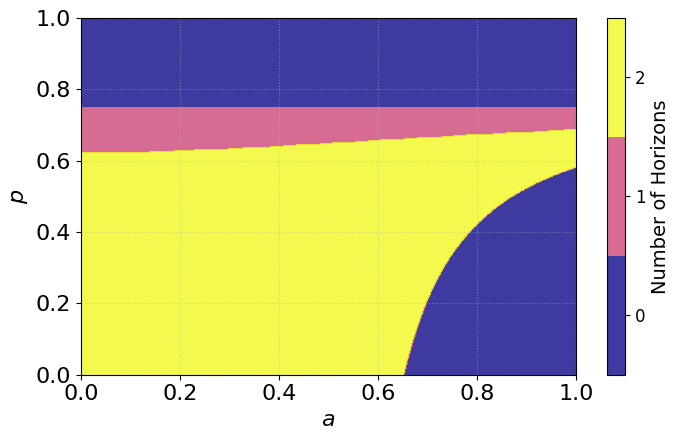} \\
    \caption{Contour plots showing the behavior of: 
        (i) $q_{\text{YM}}$ vs $a$, for $\alpha=-0.1$ and $p=0.6$ (upper figure of the first column),   (ii) $Q_{\text{YM}}$ vs $a$ for $\alpha=-0.1$ and $p=0.6$ (lower figure of the first column),(iii) $\alpha$ vs $a$, for $p=0.6$ and $q_{\text{YM}}=0.5$
        (upper figure of the second column),
        (iv) $p$ vs $a$, for $\alpha=-0.1$ and $q_{\text{YM}}=0.5$ (lower figure of the second column)}
    \label{fig:parametric_plots}
\end{figure}

\noindent The relation $Q_{YM}= \frac{2^{2p-1}}{4p-3} q^{2p}_{YM}$ tells us that the parametric values of the Yang-Mills parameter can be both positive and negative depending on $p$ values. The typical range lies in $0<p<3/4$ for which the Yang-Mills charge is positive, otherwise it is negative. In Fig.~\ref{fig:parametric_plots}, we show the contour plots of various parameters specifying the black hole spacetime. As illustrated in Fig.~\ref{fig:parametric_plots}, the number of horizons is highly sensitive to the black hole parameters. The parameter space is divided into three distinct regions: (a) the yellow region corresponds to configurations with two distinct horizons, (b) the magenta region admits only a single degenerate horizon, and (c) the purple region indicates the absence of any horizon, i.e., a naked singularity. In particular, the left panel of Fig.~\ref{fig:parametric_plots} presents a parametric plot of the Yang-Mills charge $q_\text{YM}$ as a function of the dimensionless rotation parameter $a$. From this plot, we observe that for sufficiently small values of $q_\text{YM}$, with increasing values of $a$, the horizons disappear. This suggests that there exists a critical lower bound on $q_\text{YM}$, below which the spacetime ceases to admit an event horizon for rapidly rotating configurations. The existence of two distinct horizons is maintained within the range $0.42 \leq q_\text{YM} \leq 0.58$, regardless of the value of $a$, indicating that this interval corresponds to the physically viable regime supporting smoothly defined black hole solutions with distinct inner and outer horizons. The right panel of Fig.~\ref{fig:parametric_plots} depicts the variation of another model parameter, $p$, with respect to $a$. From the plot, it is evident that for $a\lesssim 0.75$, two horizons coexist, while for $a> 0.75$, the horizons merge and eventually vanish. Thus, $a=0.75$ represents a critical threshold beyond which the spacetime no longer supports a horizon structure.

Up to this point, we have analyzed the black hole spacetime in the presence of PFDM. In the subsequent section, we turn to the computation of its geodesics.

\section{Geodesics calculation} \label{GeoCal}
The shadow and its characteristic behavior completely depend on the geometric configurations of the black hole spacetime. For example, in the spherically symmetric black hole spacetime, we have the shadow structure in a spherical shape while for the rotating spacetime, it certainly deviates from the spherical shape because of the presence of rotation parameter \cite{Hioki:2009na}. The photons coming from a distance light source and thereby deviating from their paths near black holes must follow certain geodesics. Hence, it is customary to analyze the nature of geodesics that photon follows while encountering the black hole's intense gravity. Subsequently, we need to separate out various equations regarding different coordinates. We use the variable separable methods as given by Hamilton-Jacobi and was subsequently analyzed by Carter for the axially symmetric black hole spacetimes. The general form of the Hamilton-Jacobi equation is written as

\begin{equation}
\frac{\partial S}{\partial \lambda}=-\frac{1}{2}g^{\mu \nu}\frac{\partial S}{\partial x^\mu}\frac{\partial S}{\partial x^\mu},
\label{HJ-eq}
\end{equation}
where $\lambda$ is the metric affine parameter along the geodesics. The Jacobi action $S$ is expressed in terms of a separable solution in the form
\begin{equation}
     S = \frac{1}{2}m^2\lambda - Et + L\phi + S_r(r) + S_{\theta}(\theta),
     \label{J-action}
\end{equation}
where $m$ represents the test particle mass. The quantities $E$, and $L$ correspond to energy and angular momentum pertaining the time translation and the rotational symmetries of the black hole spacetime (\ref{rotmetric}). When $m=-1$, we have a timelike particle while for null geodesics the test particle has vanishing mass $(m=0)$. In our shadow analysis, we restrict ourselves to the null geodesics. \\

\begin{figure}[H]
    \centering
    \includegraphics[width=0.45\linewidth]{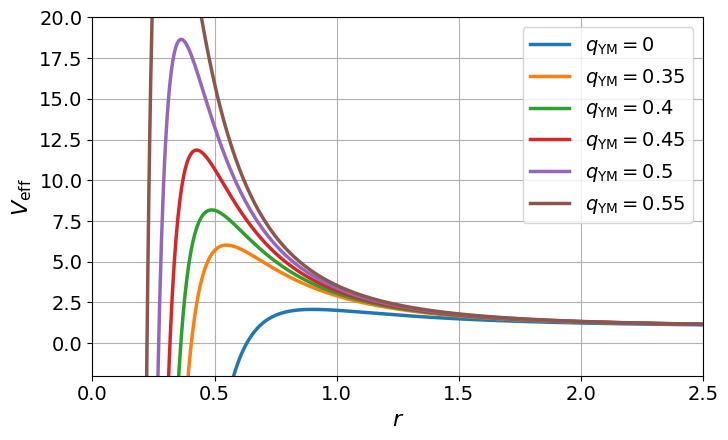}
    \includegraphics[width=0.45\linewidth]{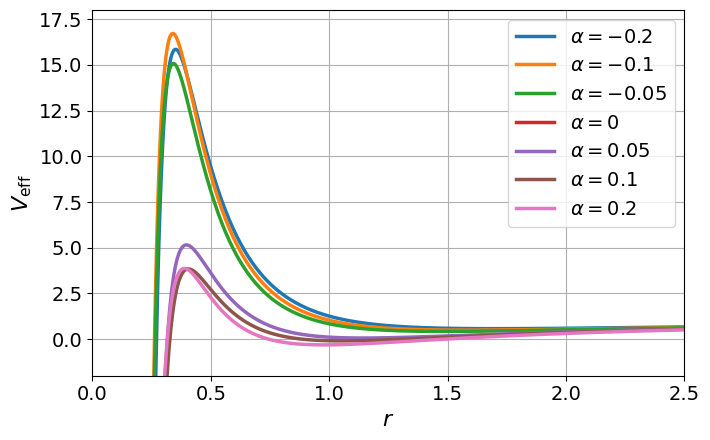}
    \caption{(i) Plot for potential for different value of $q_{YM}$ (ii) Plot for potential for different value of $\alpha$}
    \label{fig:effetive-pot}
\end{figure}

Utilizing Eqs.~(\ref{HJ-eq}) and (\ref{J-action}), and the metric (\ref{rotmetric}), we get the null geodesics equations of the test photon around the Yang-Mills field-inspired charged black hole in PFDM background as follows

\begin{eqnarray}
\rho^2\left(\frac{dt}{d\lambda}\right) &=& \frac{(r^2 +a^2)P}{\Delta} - a(aE\sin^2\theta - L),\\
\rho^2\left(\frac{dr}{d\lambda}\right)&=&\sqrt{\mathcal{R}(r)},\\
\rho^2\left(\frac{d\theta}{d\lambda}\right) &=&\sqrt{\Theta(\theta)},\\
 \rho^2\left(\frac{d\phi}{d\lambda}\right)&=&\frac{aP}{\Delta} - \left(aE - \frac{L}{\sin^2\theta}\right),
\end{eqnarray}
where the functional form of the quantities $\mathcal{R}$, ${\Theta}$, $P$ are written as
\begin{eqnarray}
&&\mathcal{R}(r)=\left[E(r^2 + a^2) - aL\right]^2 - \Delta\left[(L - aE)^2 +\mathcal{K}\right],\\
&&\Theta(\theta) =\mathcal{K}- \left[\frac{L^2}{\sin^2\theta} - a^2E^2\cos^2\theta\right],\\
&&P = E(r^2 + a^2) - aL,
\end{eqnarray}
where $\mathcal{K}$ is the Carter constant and has dimension of energy-squared. For shadow formation and its subsequent detection we need to analyze the unstable photon sphere radius. The condition for unstable circular orbit is determined from the relation
\begin{equation}
  \left(\frac{dr}{d\lambda}\right)^2 +V_{eff}(r)=0
\end{equation}
 where the effective potential $V_{eff}$ for photon is obtained in the functional form as 
 \begin{equation}
     V_{eff}(r) = \frac{\left(E\left(r^2 + a^2\right) - aL\right)^2}{\rho^4} - \frac{\Delta}{\rho^4}\left[\left(L - aE\right)^2+\mathcal{K}\right]
 \end{equation}
Fig.~\ref{fig:effetive-pot} shows the variation of effective potential with respect the radial coordinate for different set of values of the Yang-Mills charge (left) and PFDM parameter (right). We observe that with increasing values of the Yang-Mills charge, the peak of the effective potential increase significantly. This reflects the fact that the formation of unstable photon orbits is enhanced if we go on increasing the values of Yang-Mills charge. 
In contrast, when the PFDM parameter is increased, the peak exhibits a decreasing behavior. Therefore, we conclude that larger values of the Yang--Mills charge together with smaller values of the PFDM parameter are more conducive to the formation of a dent in the black hole shadow. Physically, these trends arise because both the Yang-Mills charge and the PFDM parameter modify the effective potential governing null geodesics, thereby changing the location and stability of the photon sphere responsible for shadow formation.

 \section{Black Hole Shadow}\label{shadow} 
This section is devoted to the analysis of the shadow properties of black holes. In particular, we introduce two impact parameters, which serve to characterize the corresponding shadow structure, namely,
\begin{eqnarray}
     \xi=\frac{L}{E} \\
     \eta=\frac{\mathcal{K}}{E^2}
 \end{eqnarray}
 These are the constant quantities defining the geodesics. The formation of black hole shadow is possible only when we have $\mathcal{R}(r)\geq0$ and $\mathrm{\Theta}$$(\theta)\geq0$. The choice of radial and angular velocity is either positive or negative and can be chosen independently. When one set $\mathcal{R}(r)=0$ or $\Theta(\theta)=0$, we get the turning points for the photon motion.
Accordingly, the effective potential is rewritten as
 \begin{equation}
     V_{eff}(r) = E^2\frac{\left(\left(r^2 + a^2\right) - a\xi\right)^2 - \Delta\left[\left(\xi - a\right)^2 + \eta\right]}{\rho^4}
 \end{equation}
 We define another quantity $\mathcal{V}_{eff}(r)={V_{eff}}/{E^2}$ to get rid of the quantity $E^2$ in the expression of $V_{eff}(r)$. Now, we write the mathematical form of the boundary formation of the black hole shadows which are determined through the conditions
 \begin{equation}
\mathcal{V}_{eff}(r)=\mathcal{V}^\prime_{eff}(r)=0
 \end{equation}
When light rays is passing by the black hole spacetime, it forms three types of the photon orbits, viz., it gets captured, scatters
to infinity, or form the bound orbits. When the photon gets constrained on a sphere of constant radius, we get $\dot{r}=0$ and $\ddot{r}\geq 0$ and correspondingly we have the spherical photon orbits. At this value of the photon orbit, we have the extrema of the effective potential at the unstable photon sphere with radius $r=r_p$, where $r_p$ is the radius of the photon orbit. These orbits form a $2$D dark region near the black holes called the shadow of the black hole. When solved for $\eta$ and $\xi$, we have them in terms of $r_p$ the following forms 
\begin{eqnarray}
\eta_{\text{crit}}&=& \frac{r_p^3 \left(8 a^2 f'(r_p) - r_p \left(r_p f'(r_p) - 2 f(r_p)\right)^2 \right)}{a^2 \left(r_p f'(r_p) + 2 f(r_p)\right)^2}\\
\xi_{\text{crit}}&=& \frac{(r_p^2 + a^2)\left(r_p f'(r_p) + 2 f(r_p)\right) - 4\left(r_p^2 f(r_p) + a^2\right)}{a\left(r_p f'(r_p) + 2 f(r_p)\right)}
\end{eqnarray}

In the above expressions, ${}^\prime$ denotes differentiation with respect to the radial coordinate, and $f(r)$ refers to the metric function of the static seed spacetime defined in Eq.~(\ref{metricf}), which relates to the rotating spacetime metric function $\Delta(r)$ via $\Delta(r) = r^2 f(r) + a^2$. It is to be mentioned here that in the limit when $\alpha=0$, and $q_{\text{YM}}$, the above expressions would reduce to the corresponding expressions for the Kerr-Newman black holes. \\
The shadow radius is obtained by putting the above quantities together as follows
\begin{equation}
\eta_\text{crit} +\xi^2_\text{crit} = R_s^2, 
\end{equation}
where $R_s$ is the shadow radius of the black hole, and is expressible in terms of the metric function and the black hole parameters, via,
\begin{eqnarray}
    \label{radius}
R_s^2=2r_p^2+a^2+\frac{8\Delta(r_p)\left[2-\left(r_pf'(r_p)+2 f(r_p)\right)\right]}{\left(r_pf'(r_p)+2 f(r_p)\right)^2}    
\end{eqnarray}
\begin{figure}
    \centering
    \includegraphics[width=0.39\linewidth]{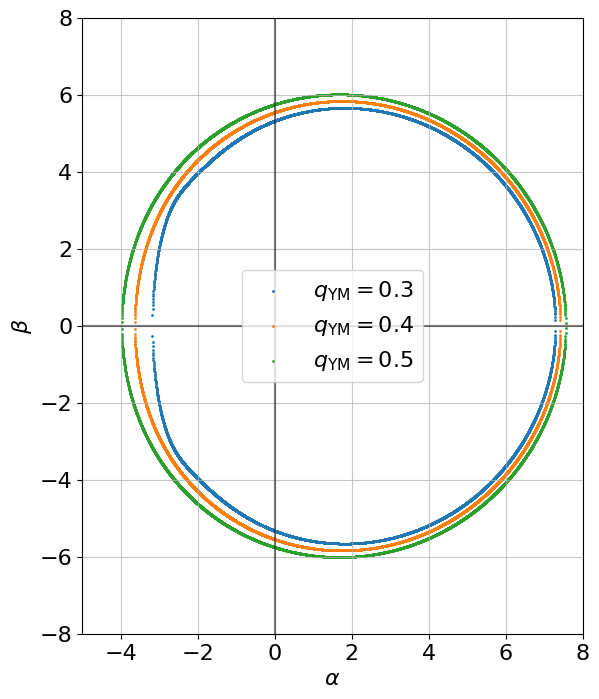}
    \includegraphics[width=0.43\linewidth]{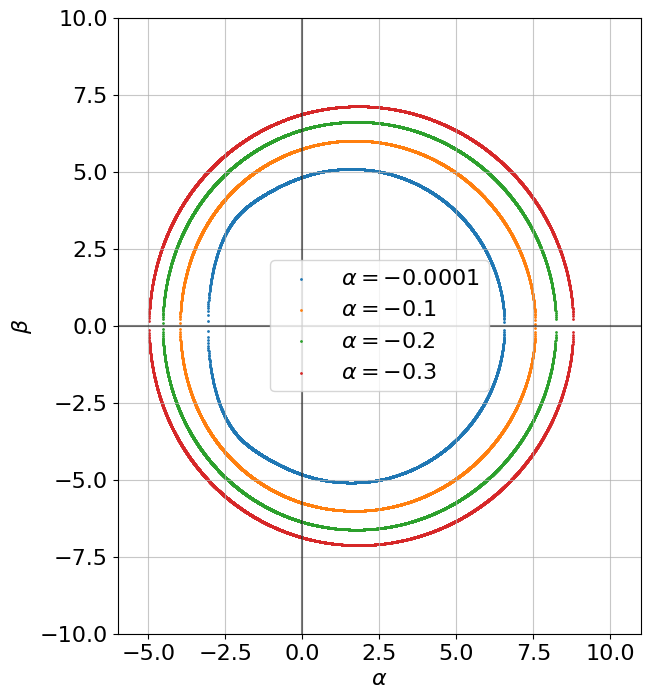}
    \includegraphics[width=0.40\linewidth]{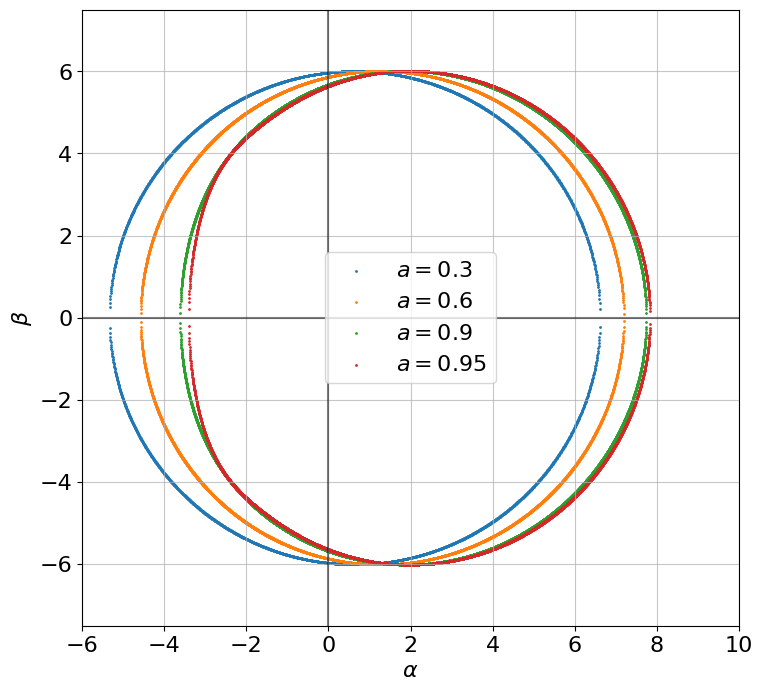}
    \includegraphics[width=0.40\linewidth]{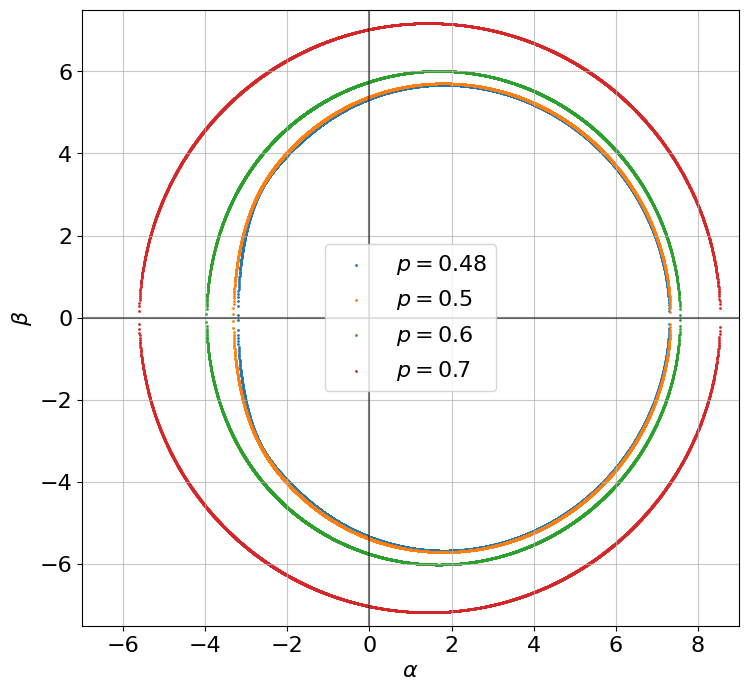}
    {\caption{Parametric plots of the charged rotating black holes shadow in Yang-Mills theory endowed with a PFDM environment--(i) for different value of $q_{\text{YM}}$ with $p=0.6$, $a=0.8$, and $\alpha=-0.1$ (ii) for different value of the PFDM parameter $\alpha$ with $p=0.6$, $a=0.8$, and $q_\text{YM}=0.5$ (iii) for different value of spin parameter $a$ with  $p=0.6$, and $q_\text{YM}=0.5$, and (iv) for various values of the parameter $p$ with $a=0.8$, $\alpha=-0.1$, and $q_\text{YM}=0.5$. (Note: where axis labels in the plots display $\alpha$ and $\beta$, they denote the celestial coordinates $X$ and $Y$ defined in the text, not the PFDM parameter).}}
    \label{fig:shadows_plot}
\end{figure}
The shadow radius of the Kerr-Newman black holes are obtained as a limiting cases when $\alpha=0,\;\text{and}\;Q_{\text{YM}}=0$, which in addition, gives the expression for Kerr black hole when $Q=0$. For non-rotating case we have to substitute $a=0$, and the expression (\ref{radius}) describes a perfectly circular shape of the black hole shadow.\\
The geometric structure of the shadow depends on the various parameters characterizing black holes. From the shadow structure, we can estimate and subsequently measure the various observable parameters like the spin, mass, charge and other parameters invoked in defining the black holes. The distant observer is at angular position $\theta_0$ subtended with the rotation axis of the black hole. We define the celestial coordinates ($X,Y$), in order to get the shadow boundary of the observer's sky (note that we explicitly use $X$ and $Y$ instead of the traditional $\alpha$ and $\beta$ to avoid any confusion with the PFDM parameter $\alpha$). These coordinates are the apparent angular distances measured from the observer's line of sight in parallel and perpendicular, respectively. In this set up the observer is situated at the distant when $r_0\to\infty$. Therefore, the coordinates $(X,Y)$ are the celestial coordinates, which are obtained when we have a stereographic projection of the black hole shadow on the observer's sky. Hence the boundary coordinates of the black hole shadow are defined to be 

\begin{eqnarray}
    \label{celestial}
    X&=&\lim_{r_0\to\infty}\left(-r_0^2\sin\theta_0\frac{d\phi}{dr}\right)\nonumber\\
     Y&=&\lim_{r_0\to\infty}\left(r_0^2\frac{d\theta}{dr}\right)
\end{eqnarray}

where $r_0$ is the distance between the observer and the black hole. For an asymptotic observer, Eq~(\ref{celestial}) reduces to

\begin{eqnarray}
    \label{celestial_01}
 X&=&-\xi_{\text{crit}}\csc\theta_0\nonumber\\
     Y&=&\pm\sqrt{\eta_{\text{crit}}+a^2\cos^2\theta_0-\xi_{\text{crit}}^2\cot^2\theta_0}
\end{eqnarray}
Putting together the expressions (\ref{celestial_01}) of shadow boundary coordinates, we have 
\begin{eqnarray}
    \label{shadow radius}
    X^2+Y^2=\eta_{\text{crit}}+\xi_{\text{crit}}^2+a^2\cos^2\theta_0
\end{eqnarray}

Fig.~\ref{fig:shadows_plot} represents the parametric plots depicting the shadow of the charged rotating black holes in Yang-Mills theory immersed in PFDM background. The shadows of our concerned black holes are depicted with respect to different parametric set of values. In the upper panel of Fig.~\ref{fig:shadows_plot}(left) , we take a parametric set of values of $q_{\text{YM}}$ for fixed values of $\alpha$, $a$ and $Q$. We can see that as we increase the values of $q_{\text{YM}}$ there is an increase in the shadow size. The appearance of a more visible dent is observed when $q_{\text{YM}}$ has the smaller values, thus contributing significantly to the circularity deviation in the shadow shape. In the upper panel of Fig.\ref{fig:shadows_plot} (right), as we increase the values of the PFDM $\alpha$, there is a decrease in the shadow size and its shape deviates remarkably. In the lower panel of Fig.~\ref{fig:shadows_plot}(left), the appearance of the dent with increasing values of the spin parameter $a$ is quite obvious. As we increase the values of $a$ the dent becomes clear as it is expected for any rotating black hole spacetime. We also demonstrate the variation in the shadow shape with different values of the parameter $p$ as can be seen from the Fig.~\ref{fig:shadows_plot}(right). These results demonstrate that the shadow observables are sensitive to both the intrinsic black hole parameters and the surrounding dark matter environment, suggesting that precise shadow measurements may provide complementary information for constraining non-Kerr black hole models.
\section{Observables and black hole parameter estimation}\label{ObBHpara}
In this section, we focus on constraining the parametric space of Yang--Mills--modified charged black holes embedded in a PFDM environment. While the EHT analysis of the M$87$ supermassive black hole primarily adopted the Kerr spacetime as its fiducial model, it is important to emphasize that the collaboration did not exclude black hole solutions arising from modified gravity theories as viable candidates for their target spacetimes. Motivated by this perspective, we utilize the EHT observational bounds to estimate constraints on the characteristic parameters of the Yang--Mills--PFDM black hole, namely the Yang--Mills charge $q_{\text{YM}}$, the Maxwell charge $Q$, the PFDM parameter $\alpha$, and the dimensionless spin parameter $a$.

The first-order corrections to the circularity of the black hole shadow appear in the Kerr background once the spin parameter is introduced. In contrast, for black holes arising in modified gravity frameworks, additional distortions are induced through the variation of other coupling parameters characterizing the spacetime. This necessitates the computation of shadow observables, which can then be employed to place bounds on the underlying black hole parameters in these non-Kerr geometries. 

As a representative approach, Hioki and Maeda \cite{Hioki:2009na} proposed a method to determine the shadow radius $R_s$ and the distortion parameter $\delta_s$ by imposing certain symmetry assumptions on the shadow contour. However, this method is applicable only when such symmetry conditions are satisfied. In more general settings, where the shadow may exhibit irregular or asymmetric deformations, more robust and model-independent techniques are required. Such methods rely on the full shadow morphology and allow for parameter estimation of rotating black holes in a wide class of modified gravity theories, as discussed in \cite{Johannsen:2013vgc,Tsukamoto:2014tja,Wang:2017hjl,Tsupko:2017rdo,Kumar:2018ple}.

As is mentioned earlier, the EHT observations can constrain the physical quantities specially the black hole mass and other related parameters. However, the EHT does not include the measurement of the angular momentum parameters in its early operations. Nevertheless, the estimations of the mass parameters of the black holes M$87^*$ and Sgr$A^*$ are included in their measurements. Following the techniques of determining the shape and size of the black holes as proposed in \cite{Kumar:2018ple}, we can characterize the shadow of the black holes without any approximation. The actual shadow area of the rotating black hole in any modified gravity theories could be written as\cite{Kumar:2018ple}
\begin{equation*}
    A = 2\int{Y(r_p)dX(r_p)} \\
      =2 \int_{r_p^-}^{r_p^+}\left(Y(r_p)\frac{dX(r_p)}{dr_p}\right) dr_p,
\end{equation*}
and its oblateness by
\begin{equation}
    D=\frac{X_r-X_l}{Y_t-Y_b}
\end{equation}
where $X_l,\;\text{and}\;X_r$ represent the left and right of the shadow boundary when expressed in the celestial coordinates $(X,Y)$ and in such case we set $Y(r_p)=0$. On the other hand, $Y_t$ and $Y_b$ correspond to the top and bottom ends of the shadow boundary in the same celestial coordinates $(X,Y)$ and in this case we have $Y^\prime(r_p)=0$. It is worth mentioning here that for a spherically symmetric black holes in any theory of gravity, we have $D=1$, while for Kerr case it lies in the range, i.e., $\sqrt{3}/2\leq D\leq1$ \cite{Tsupko:2017rdo}. Whenever we have a rotating axisymmetric spacetimes, the parameter $D$ never equals to unity and always has values below or above unity \cite{Johannsen:2010ru}. 
\begin{figure}[H]
    \centering
    \includegraphics[width=0.45\linewidth]{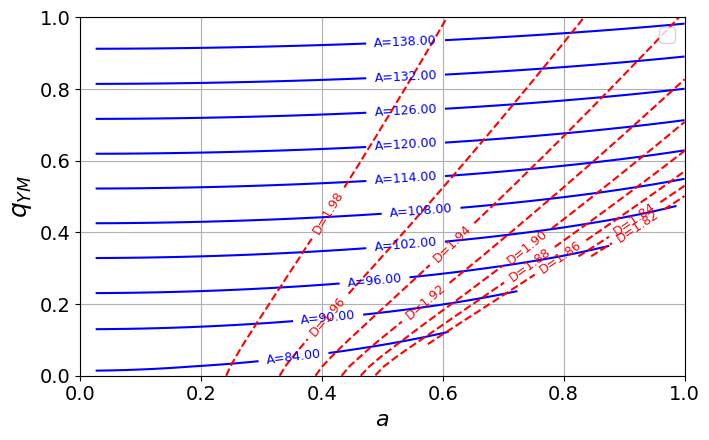}
    \includegraphics[width=0.45\linewidth]{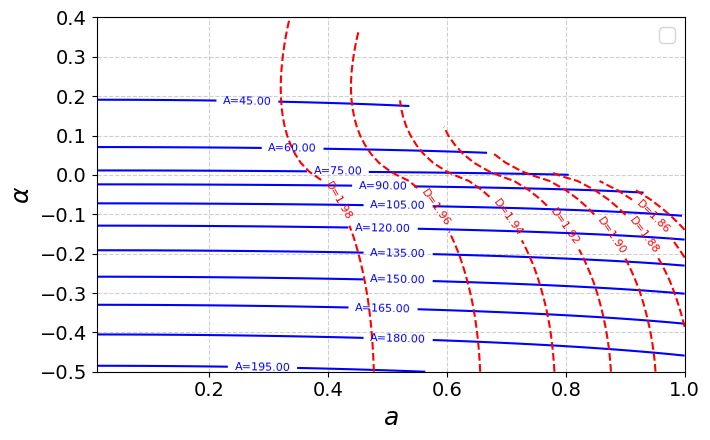}\\
    \includegraphics[width=0.45\linewidth]{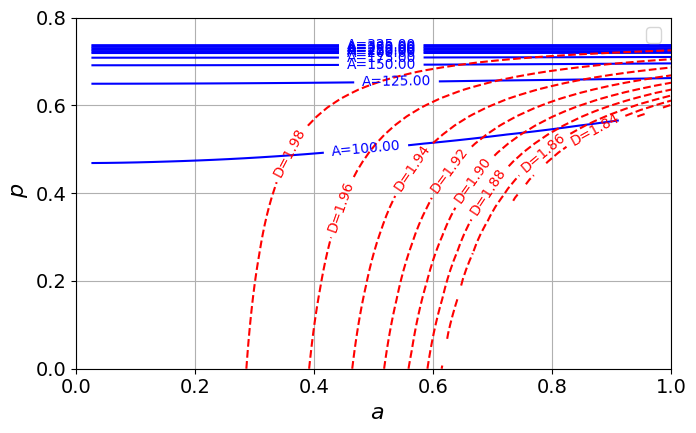}
    \caption{Contour plots of shadow area observables $A$ and oblateness $D$ in the parameter space (i) $(a,q_{\text{YM}})$ (ii) $(a,\alpha)$ (iii) $(a,p)$. In each curve we designate the corresponding values of the area observable $A$ (blue solid lines) and oblateness $D$.}
    \label{fig:contour_Area}
\end{figure}
We depict in Fig.~\ref{fig:contour_Area}, the contour plot of the shadow area and oblateness of the black hole of our interest. On the upper panel of the Fig.~\ref{fig:contour_Area}, we depict the area observable ($A$) and oblateness ($D$) in the parameter spaces $(a,q_{\text{YM}})$ (left), the parameter space $(a,\alpha)$ (right). On the lower panel, we plot the contours in the parameter space $(a,p)$. The blue solid lines in each figure represent the area observable whereas the red dashed lines correspond to the oblateness. In the parameter spaces of $(a,q_{\text{YM}})$, $(a,\alpha)$, and $(a,p)$, we have the degeneracies of the quantities  $A$ and $D$, provided that we can get more than two combinations of the various black hole parameters in a pairwise fashion including $(a,q_{\text{YM}})$, $(a,\alpha)$, and $(a,p)$, as demonstrated in Fig.~\ref{fig:contour_Area}. On the other hand, for a given observable $A$ or $D$, one can establish possible correlations between various parameters  such as $(a,q_{\text{YM}})$ or $(a,\alpha)$ or $(a,p)$.
 Since $A$ and $D$ intersect at a fixed point, this leads to conclude that the observables $A$ and/or $D$ are also having non-degenerate values in various parametric spaces, if any of these pairs are kept fixed.

\section{Constraints from EHT observations of M87 and SgrA\texorpdfstring{$^*$}{*}}\label{sag}
The shadow analysis using the results from EHT observations and modeling rotating black holes in various modified gravity theories always gives some hints regarding the possible constraints on the parameters defining the black holes. After the first ever shadow detections \cite{EventHorizonTelescope:2019dse} of supermassive black holes M$87$ via EHT, there is a surge of interest to test other rotating black holes in of the EHT observations become unprecedentedly useful and essential tool to test various black holes in modified gravity. For instance, we could determine the mass of the rotating space-times using  the similar techniques used for M$87$ supermassive black holes. The EHT collaboration, for the first time determined $1.3$ mm image of M$87$ black hole with an angular diameter of $42\pm 3\mu$ as. The shadow images of this supermassive black hole when compared to the predicated Kerr shadow is found to be consistent. This comparison of the images using the different simulations and imaging techniques of M$87$ black holes leading the EHT collaboration to estimate the mass as $M=(6.5\pm 0.7)\times 10^9 {M_{\odot}}$. On the other hand, with the recent analysis of the shadow image of supermassive black hole SgrA$^*$, in addition to its angular diameter ($51.8\pm2.3\mu$as), the EHT collaboration also determines its shadow diameter $d_{sh}(48.7\pm7\mu$ as)\cite{EHT2022_Paper1, EHT2022_Paper2, EHT2022_Paper3, EHT2022_Paper4, EHT2022_Paper5}. The measured values of the mass and the shadow diameter, and the circularity deviation parameter $\Delta C$ are obtained in accordance with the shadow of the Kerr black hole as a seed metric.  Although the EHT collaboration use the Kerr black hole hypothesis to image shadow of the supermassive black holes M$87$ and SgrA$^*$, because of the uncertainties in determining the rotation parameter we cannot ignore other rotating black holes coming from various modified gravity theories \cite{Kumar:2018ple,Cunha:2019ikd,Kumar:2019ohr,Vagnozzi:2019apd}. Rather, we can put constraints on the spin parameter as well as the other parameters defining the Kerr-modified rotating black holes. In such modeling we put the same angular diameter $\theta_d$ and the circularity deviation $\Delta C$ in our model and infer information of important parameters apart from rotation parameter defining our concerned spacetimes. In addition to $\theta_d$ and $\Delta C$, we can also have parametric constraints through other quantities, namely, the shadow deviation parameter $\delta$ and the shadow diameter $d_{sh}$. 

\begin{figure}[H]
    \centering
    \includegraphics[width=0.45\linewidth]{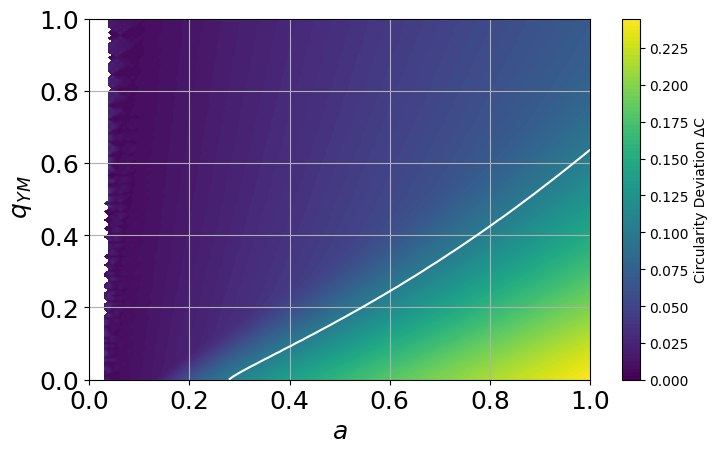}
    \includegraphics[width=0.45\linewidth]{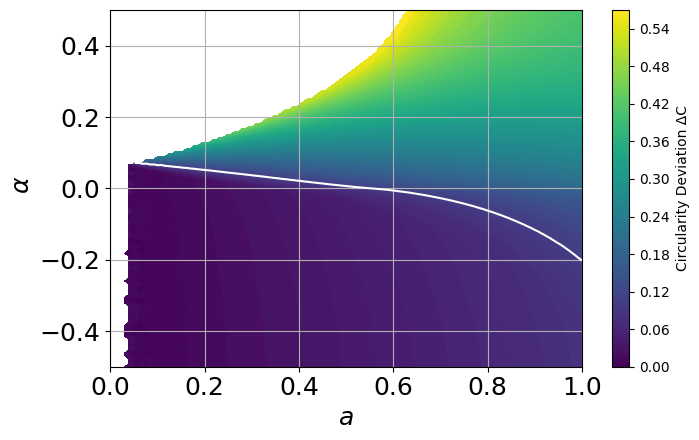}\\
    \caption{(i) The Circularity deviation $\Delta C$ for M$87$ supermassive black hole in the parameter space $(a,q_\text{YM})$ for $p=0.6$ and $\alpha=-0.1$ (left) and (ii) in the parameter space $(a,\alpha)$ for $p=0.6$ $q_\text{YM}=0.5$ (right)}
    \label{circularity_deviation_87}
\end{figure}

Hence, in this article, we model Yang-Mills inspired charged PFDM rotating spacetime with the supermassive black holes M$87$ and SgrA$^*$ to determine bounds on various parameters. Hence, for the SgrA$^*$ at the center of our Milky Way galaxy, and M$87$ at the nearby galactic center, we consider the central values for the masses $M=4.3\times 10^6 M_{\odot}$ and $M=6.5 \times 10^9 M_{\odot}$, respectively, and the observer distances to be $d=8.35$~Kpc, and $d=16.8$~Mpc, respectively. Note that the uncertainty in the mass measurement of M87* is relatively significant and could have a non-negligible impact on the resulting constraints; while we focus on the central mass value in the current parameter estimation, a more robust future analysis should rigorously incorporate this observational uncertainty. Furthermore, regarding the observer inclination angles used in our analysis, we consider $\theta_0=17^\circ$ for M87*, which is physically motivated by the orientation of its large-scale jet \cite{EventHorizonTelescope:2019dse}, and we additionally consider $\theta_0=90^\circ$ (equatorial plane) as a limiting case to illustrate the maximal rotational distortion of the shadow. For SgrA$^*$, recent EHT results favor a relatively lower inclination angle (often inferred to be $\lesssim 50^\circ$) \cite{EventHorizonTelescope:2022wkp,EventHorizonTelescope:2022xqj}; accordingly, we adopt an inclination of $\theta_0=50^\circ$ as a representative realistic value, alongside the maximal $\theta_0=90^\circ$ case to ensure a comprehensive parameter space study. \\
Before we define the circularity deviation, $\Delta C$, we need to analyze the average shadow radius of the rotating black holes by the boundary coordinates $(R(\varphi),\varphi)$. We also need to identify the center of the circle as represented by the coordinates $(X_c,Y_c)$, where $X_c=\left(X_r-X_l\right)/2$ and $Y_c=0$, admitting a intrinsic reflection symmetry around $X$-axis. The average shadow radius is denoted by \cite{Johannsen:2010ru} 
\begin{equation}
    R(\varphi)=\frac{1}{2\pi}\int_0^{2\pi}R(\varphi)d\varphi
    \end{equation}, 

with the angle $\varphi=\arctan{\left(\frac{Y}{X-X_c}\right)}$ and $R(\varphi)=\sqrt{(X-X_c)^2+(Y-Y_c)^2}$. Once we get $\bar{R}$, we can easily determine the circularity deviation parameter 
\begin{equation}\Delta C=\frac{1}{\bar{R}}\sqrt{\int_0^{2\pi}\left(R(\varphi)-\bar{R}\right)^2d\varphi}
\end{equation}
As the definition suggests, $\Delta C$ measures possible deviations from a perfect circle. We depict in Fig.~\ref{circularity_deviation_87}, the contour plots of the circularity deviation of M$87$ black holes in the parameter spaces ($a,q_\text{YM}$) and ($a,\alpha$).

\begin{figure}[H]
    \centering
    \includegraphics[width=0.45\linewidth]{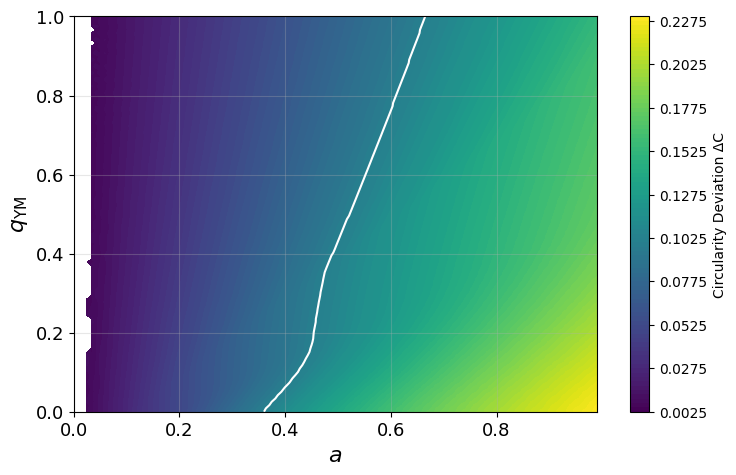}
    \includegraphics[width=0.45\linewidth]{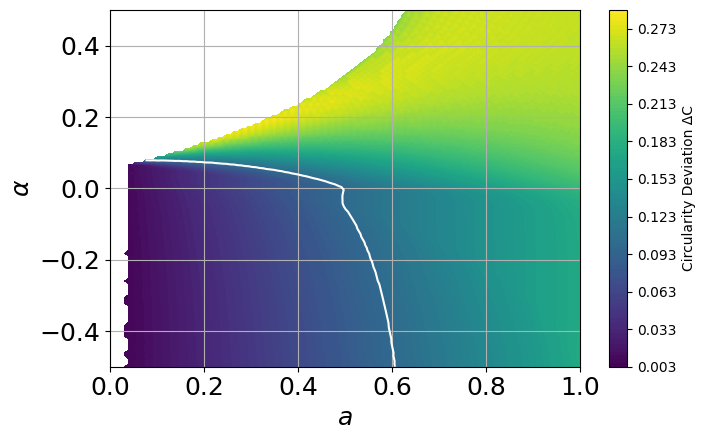}\\
    \caption{(i) The Circularity deviation $\Delta C$ for SgrA$^*$ supermassive black hole in the parameter space $(a,q_\text{YM})$ for $p=0.6$ and $\alpha=-0.1$ (left) and (ii) in the parameter space $(a,\alpha)$ for $p=0.6$ $q_\text{YM}=0.5$ (right)}
    \label{circularity_deviation}
\end{figure}

As reported earlier in the literature, these plots do not show the similar behavior of Kerr-Newman or regular black holes rather it has a unique feature reflecting the effects of Yang-Mills charge and the PFDM parameter. As reported by EHT collaboration, the upper bound on the circularity deviation for M$87$ supermassive black hole is found to be $\Delta{C}\leq0.10$ \cite{EventHorizonTelescope:2019ggy,EventHorizonTelescope:2019pgp,EventHorizonTelescope:2019dse}. Putting this bound for the Yang-Mills-inspired charged PFDM black holes, we can estimate the permissible bounds on $q_\text{YM}$ and $\alpha$. The Fig.~\ref{circularity_deviation} clearly demonstrate the interplay of the relationship between Yang-Mills charge versus spin parameter and PFDM parameter versus spin parameter. We can estimate that for our concerned black hole if we put the constraints on the circularity deviation $\Delta{C}\leq0.10$, the bounds on the Yang-Mills charge $q_\text{YM}\leq 0.7$ and the PFDM parameter $\alpha\leq 0.35$.

Here, we need to mention that the angular diameters of M87 and SgrA$^*$ as $42\pm 3\ \mu\text{as}$  and $51\pm 2.3\ \mu\text{as}$, respectively. The shadow diameter as measured by a distant observer at a distance $d$ from the black hole, is found to be \begin{equation}
    d_{sh}=2\frac{R_a}{d},\;R_a=\sqrt{A/\pi},
    \end{equation} where $R_a$ is the shadow areal radius. The shadow diameter for the Schwarzschild black hole is $6\sqrt{3}M$ and the modeled shadow diameter as determined from EHT is denoted as $\bar{d}_{\text{metric}}$ and hence the deviation is measured to be 
    \begin{equation}
    \delta=\frac{\bar{d}_{\text{metric}}}{6\sqrt{3}}-1
    \end{equation} 
    where $\bar{d}_{\text{metric}}=2R_a$. We can use these relations to put constraints on the various parameters of the rotating black holes.

\begin{figure}[H]
    \centering
    \includegraphics[width=0.45\linewidth]{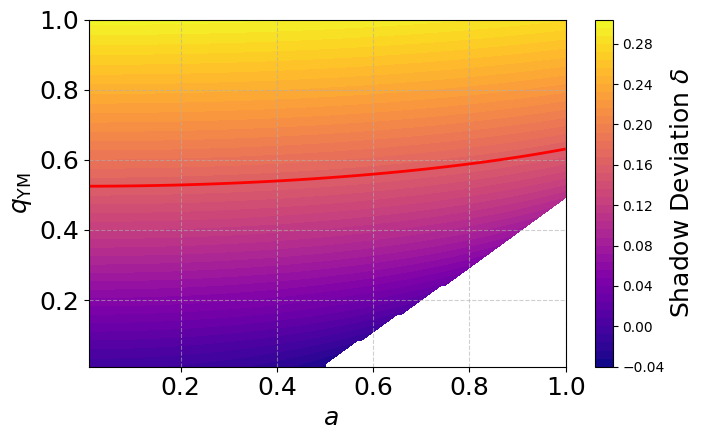}
    \includegraphics[width=0.45\linewidth]{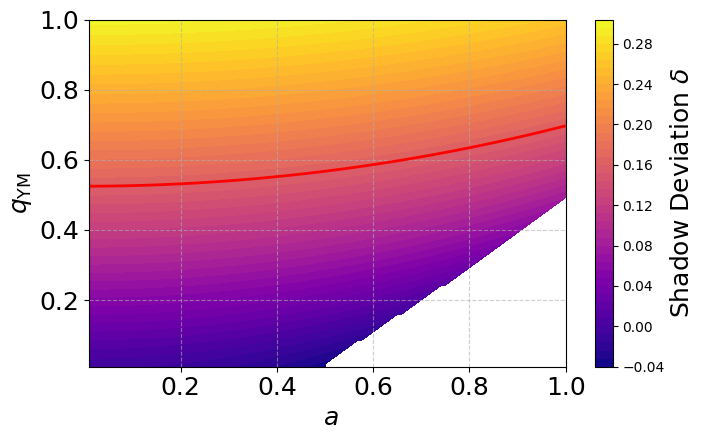}\\
    \includegraphics[width=0.45\linewidth]{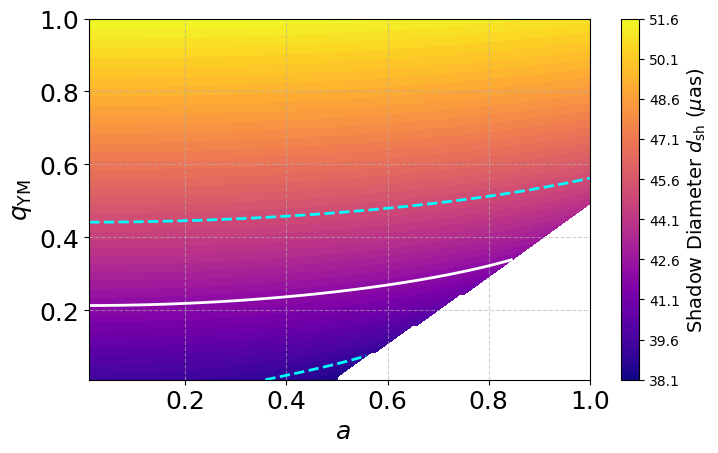}
    \includegraphics[width=0.45\linewidth]{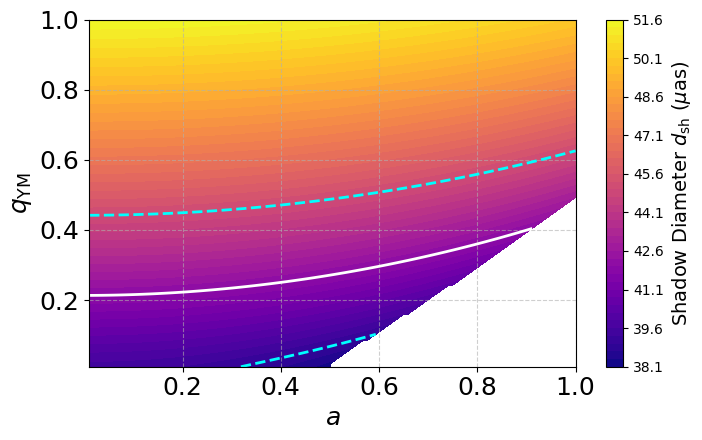}
    \caption{On the upper panel, we depict the contour plots of the shadow deviation parameter $\delta$ in the parameter spaces of (i) $(a,q_\text{YM})$ (left figure for the observing angle $\theta=90^\circ$) and (ii) $(a,q_\text{YM})$  (right figure for the observing angle $\theta=17^\circ$). On the lower panel we show the contour plots of the shadow angular diameter $d_{sh}$ for the parameter spaces (i) $(a,q_\text{YM})$ (left figure for the observing angle $\theta=90^\circ$) and (ii) ($(a,q_\text{YM})$) (right figure for the observing angle $\theta=17^\circ$)}
    \label{contours_delta_dsh}
\end{figure}

In Fig.~\ref{contours_delta_dsh}, we show the contour plots of shadow deviation parameter $\delta$ (upper panel) and the shadow angular diameter $d_{\text{sh}}$ (lower panel). It is to be mentioned that to calculate the deviation parameter we take the shadow image of Schwazrschild black hole as a perfect circle and then express how much the shadow images of our concerned black hole is deviated. As depicted in Fig~\ref{contours_delta_dsh}, the contour plot (right figure on the upper panel) of the deviation parameter is demonstrated in the parameter space $(a,q_\text{YM})$ for $\alpha = -0.1$, when modeling the Yang-Mills charged PFDM black hole as M$87$.

  We found that the parametric ranges for the spin parameter $a$ and the Yang-Mills charge $q_\text{YM}$ within $1\sigma$ bound are found to ${a}\in [0.010,1.000]$, and $q_{YM}\in [0.495, 0.654]$, respectively for the observation angle $\theta=90^\circ$. Similarly for the right figure on the upper panel the parametric ranges for $a$ and $q_\text{YM}$ are found to be ${a}\in [0.010, 1.000]$, and $q_{\text{YM}}\in [0.010, 0.628]$ respectively, for $\alpha=-0.1$ at an observation angle $\theta=17^\circ$. On the other hand, the contour plots for the shadow angular diameter $d_{sh}$ are plotted on the lower panel. From these figures the parametric ranges for $a$ and $q_\text{YM}$ within $1\sigma$ are calculated to be ${a}\in [ 0.010, 0.617]$, and $q_{\text{YM}}\in [ 0.010, 0.834]$, respectively, for $\alpha=-0.1$ and $\theta=90^\circ$ (left). For the observation angle $\theta=17^\circ$ these ranges for $a$ and $q_\text{YM}$ are derived to be ${a}\in [ 0.010, 0.559]$, and $q_{\text{YM}}\in [ 0.010, 0.767]$ for $\alpha=-0.1$. From these figures, it is observed that the shadow angular diameter puts more restricted bounds on $a$ and $q_\text{YM}$. These bounds observationally viable and could be experimented in the future observations.

\begin{figure}[H]
    \centering
    \includegraphics[width=0.45\linewidth]{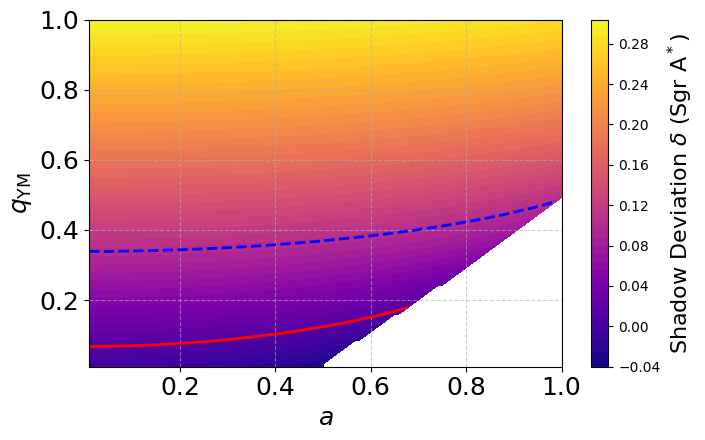}
    \includegraphics[width=0.45\linewidth]{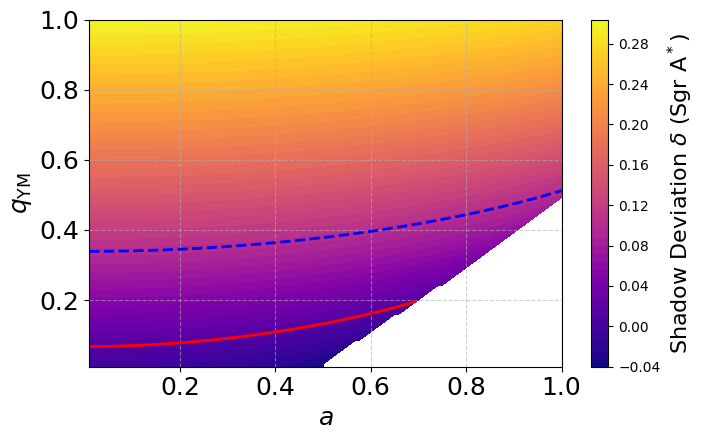}\\
    \includegraphics[width=0.45\linewidth]{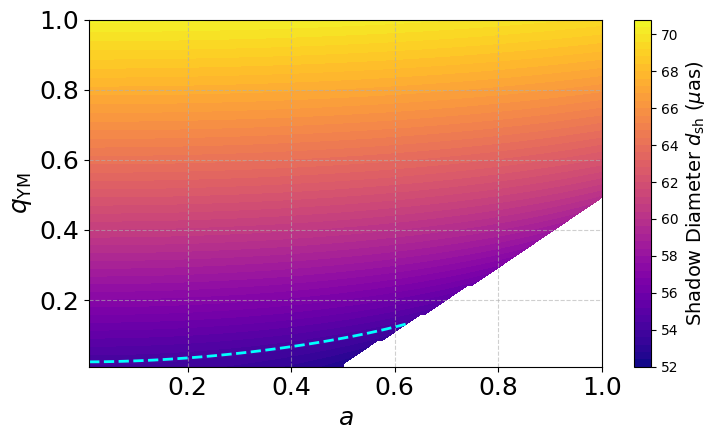}
    \includegraphics[width=0.45\linewidth]{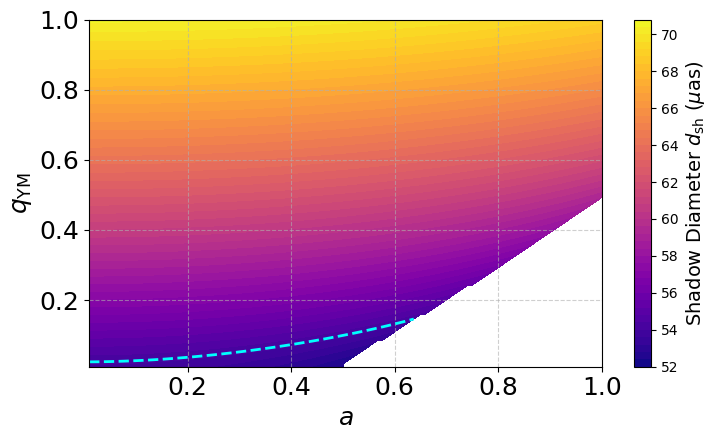}
     \caption{On the upper panel, we depict the contour plots of the shadow deviation parameter $\delta$ of SgrA$^*$ in the parameter spaces of (i) $(a,q_\text{YM})$ (left figure for the observing angle $\theta=50^\circ$) and (ii) $(a,q_\text{YM})$  (right figure for the observing angle $\theta=90^\circ$). On the lower panel we show the contour plots of the shadow angular diameter $d_{sh}$ for the parameter spaces (i) $(a,q_\text{YM})$ (left figure for the observing angle $\theta=90^\circ$) and (ii) ($(a,q_\text{YM})$) (right figure for the observing angle $\theta=90^\circ$)}
    \label{fig:SgrA1s2s}
\end{figure}

\noindent
Next we consider the observation angles $\theta_0=50^\circ,\;\text{and}\;90^\circ$ to put constraints on the parameters of YM-inspired charged PFDM black hole, modeling for SgrA$^*$ as shown in Fig~\ref{fig:SgrA1s2s}. We demonstrate that within the $1\sigma$ bounds the constraints for $\delta$ at $\alpha = -0.1$ for SgrA$^*$ on the parameters $a\in [0.010,1.000]$, and $q_{YM}\in [0.495, 0.654]$ for $\theta=50^\circ$ (left figure on the upper panel), while for the shadow angular diameter $d_{sh}$ (left figure on the lower panel) for $\alpha = -0.1$ within  $1\sigma$ bounds is found be $a\in [0.010,0.784]$, $\alpha\in [0.010,0.276]$ for the observation angle $\theta=50^\circ$,. On the other hand, the right figure on the lower panel for $\delta$ and $\alpha = -0.1$, the bounds on $a$ and $q_\text{YM}$ are estimated to be $a\in[0.010,0.784]$, $q_\text{YM}\in [0.010,0.276]$ while for $d_{sh}$ the bounds on the parameters are found to be $a\in [0.010,0.759]$, $q_\text{YM}\in [0.010, 0.251]$ at an observation angle $\theta=90^\circ$.

\begin{figure}[H]
    \centering
    \includegraphics[width=0.45\linewidth]{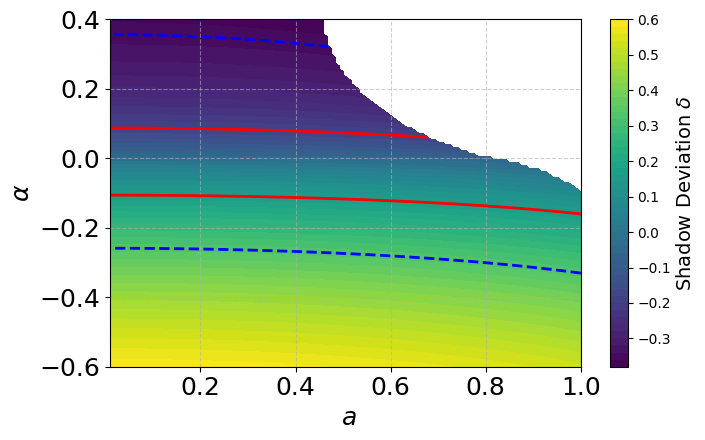}
    \includegraphics[width=0.45\linewidth]{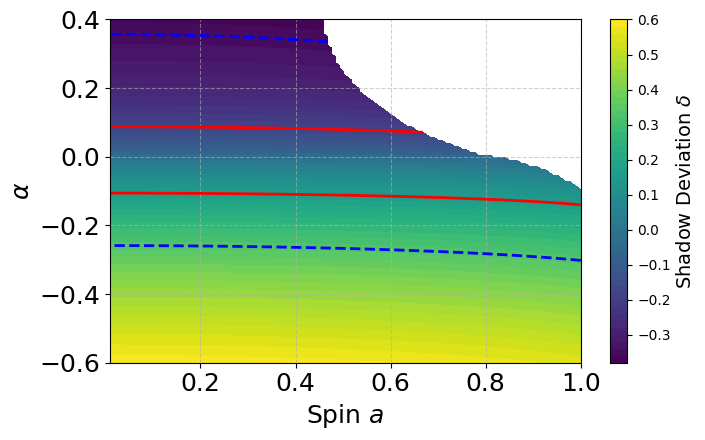}\\
    \includegraphics[width=0.45\linewidth]{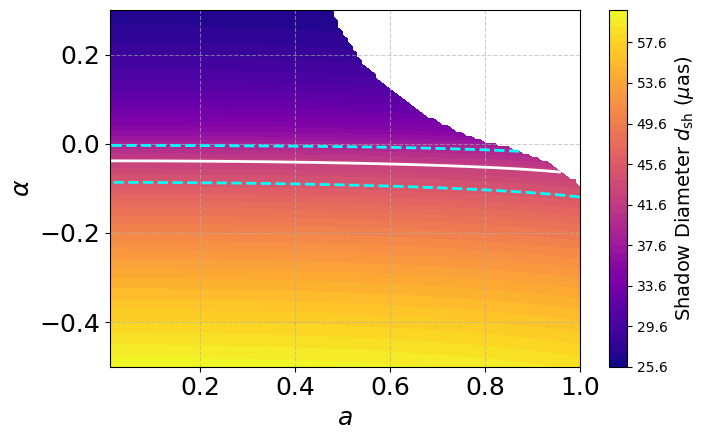}
    \includegraphics[width=0.45\linewidth]{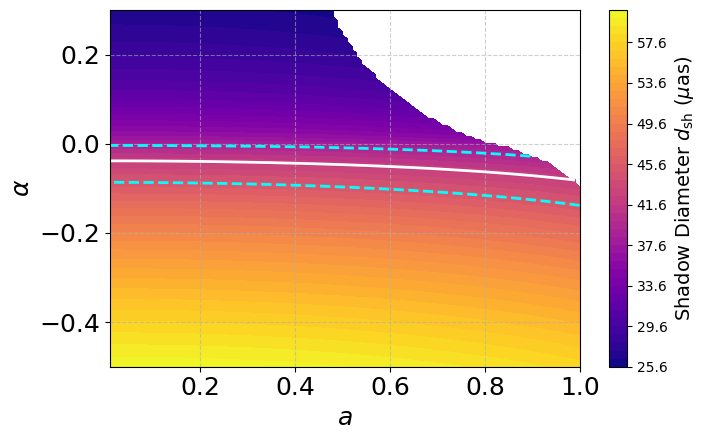}
    \caption{On the upper panel, we depict the contour plots of the shadow deviation parameter $\delta$ in the parameter spaces of (i) $(a,q_\text{YM})$ (left figure for the observing angle $\theta=90^\circ$) and (ii) $(a,q_\text{YM})$  (right figure for the observing angle $\theta=17^\circ$). On the lower panel we show the contour plots of the shadow angular diameter $d_{sh}$ for the parameter spaces (i) $(a,q_\text{YM})$ (left figure for the observing angle $\theta=90^\circ$) and (ii) ($(a,q_\text{YM})$) (right figure for the observing angle $\theta=17^\circ$)}
    \label{fig:M87-alpha}
\end{figure}

In Fig.~\ref{fig:M87-alpha}, we show the contour plots for the deviation parameter $\delta$ for $q_{YM} = 0.5$ of M$87$ black holes at the observation angles $\theta=17^\circ$ and $90^\circ$ within $1\sigma$. The parametric bounds on the parameters ${a}\in [0.010, 1.000]$, and $\alpha\in [-0.155,0.081]$ and within $2\sigma$ is ${a}\in [0.010, 1.000]$, $\alpha\in [-0.323,0.350]$ at an observation angle $\theta=17^\circ$ for the left figures on the upper panel. Similarly for the right figure on the upper panel, for $q_\text{YM} = 0.5$ the bounds on the parameters for $1\sigma$ are estimated to be ${a}\in [0.010, 1.000]$, $\alpha\in [-0.135,0.087]$ and within $2\sigma$ is ${a}\in [0.010, 1.000]$, $\alpha\in [ -0.297,0.349]$ for $\theta=90^\circ$. For the shadow angular diameter $d_{sh}$ for $q_{YM} = 0.5$, the constraints within $1\sigma$ bounds are expected to be ${a}\in [0.010, 1.000]$, and $\alpha\in [-0.136,-0.007]$ and within $2\sigma$ these bounds are ${a}\in [0.010, 1.000]$, $\alpha\in [ -0.201,0.025]$ for $\theta=17^\circ$ (left figure on the lower panel). On the right figure of lower panel for $q_{YM} = 0.5$ within $1\sigma$, the parameters are bounded to be in the ranges ${a}\in [0.010, 1.000]$, and $\alpha\in [-0.112,-0.007]$ and within $2\sigma$ bounds they are expected to be in the ranges ${a}\in [0.010, 1.000]$, $\alpha\in [ -0.177, 0.025]$ for $\theta=90^\circ$ 

\begin{figure}[H]
    \centering
    \includegraphics[width=0.45\linewidth]{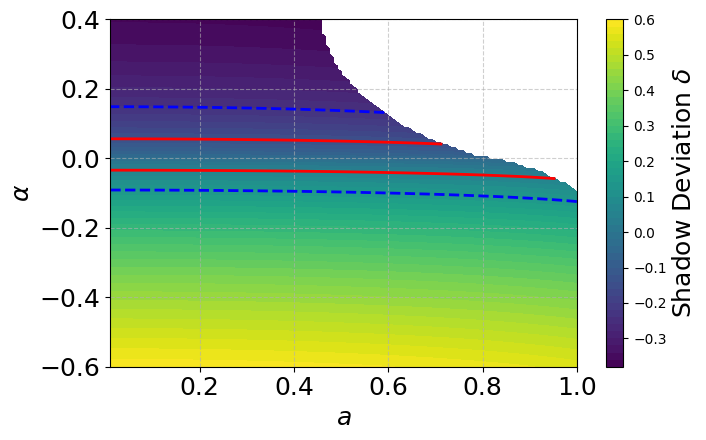}
    \includegraphics[width=0.45\linewidth]{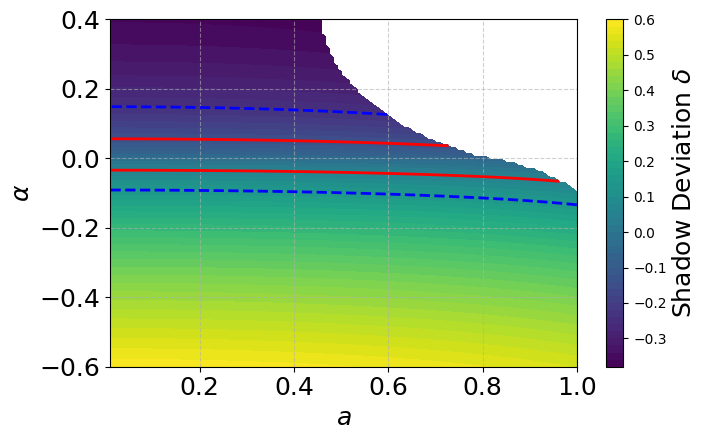}\\
    \includegraphics[width=0.45\linewidth]{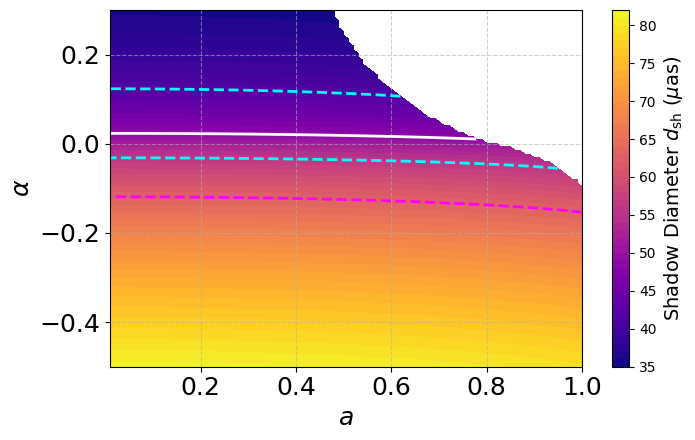}
    \includegraphics[width=0.45\linewidth]{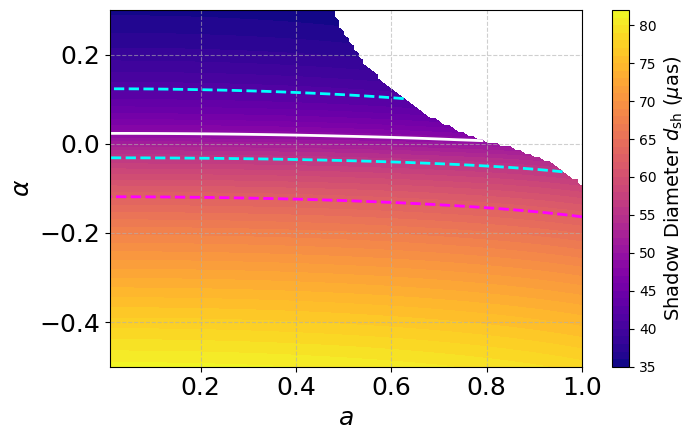}
    \caption{On the upper panel, we depict the contour plots of the shadow deviation parameter $\delta$ in the parameter spaces of (i) $(a,q_\text{YM})$ (left figure for the observing angle $\theta=90^\circ$) and (ii) $(a,q_\text{YM})$  (right figure for the observing angle $\theta=50^\circ$). On the lower panel we show the contour plots of the shadow angular diameter $d_{sh}$ for the parameter spaces (i) $(a,q_\text{YM})$ (left figure for the observing angle $\theta=90^\circ$) and (ii) ($(a,q_\text{YM})$) (right figure for the observing angle $\theta=50^\circ$)}
    \label{fig:SgrA-alpha}
\end{figure}

Finally, In Fig.~\ref{fig:SgrA-alpha}, we show the contour plots for the deviation parameter $\delta$ for $q_{YM} = 0.5$ of SgrA$^*$ black holes at the observation angles $\theta=50^\circ$ and $90^\circ$ within $1\sigma$. The parametric constraints on the parameters for $q_\text{YM} = 0.5$ within $1\sigma$ bounds are $a\in[0.010, 0.958]$, $\alpha\in[-0.062, 0.055]$ and within $2\sigma$ is $a\in [0.010, 1.000]$, $\alpha\in [-0.129, 0.148]$ for the observation angle $\theta=50^\circ$ (left figure on the upper panel). On the right figure of the upper panel for $q_{ym} = 0.5$, these bounds within $1\sigma$ level are expected to be in the ranges for $a\in [0.0100, 0.9501]$, and $\alpha/M\in [-0.0538, 0.0555]$ and within $\sigma_2$ bounds they are in the range $a\in [0.0100, 1.0000]$, and $\alpha\in [-0.1210, 0.1479]$ for the observation angle $\theta=90^\circ$. Similarly, on the left figure of lower panel we show the contour plots for shadow angular diameter $d_{sh}$ for $q_\text{YM} = 0.5$ within $1\sigma$ bounds and find the parametric ranges are $a\in [0.010, 0.950]$, $\alpha\in  [-0.056, 0.122]$ and within $2\sigma$ these bounds are estimated to be $a\in [0.010, 1.000]$, and $\alpha \in[-0.161, 0.300]$ for $\theta=50^\circ$. In the right figure on the lower panel we find for the contour plots of $d_{\text{sh}}$ for $q_{\text{YM}}= 0.5$, the bounds on the parameters within $1\sigma$ as $a\in[0.010, 0.940]$ , $\alpha\in[-0.047, 0.122]$ and within $2\sigma$ these ranges are expected as $a\in[0.010, 1.000]$, $\alpha\in[-0.153, 0.300]$ at the observation angle $\theta=90^\circ$.

\section{Energy Emission Rate}\label{eer}
We discuss about the energy emission rate in connection with the black hole shadow radius, and the horizon temperature. For an asymptotic observer located at spatial infinity ($r_0\to\infty$), the black hole shadow radius has a similar behavior of the geometric cross-section in the eikonal limit. It is argued that for the observer at asymptotic infinity the absorption cross-section has a direct correspondence with the black hole shadow radius. In this sense, the geometric cross-section has the similar form of the absorption cross-section and supposed to oscillate near $\sigma_{\text{lim}}$, a constant limiting value of the absorption cross-section such that 
\begin{equation}
    \sigma_{\text{lim}}\approx \pi R_s^2,
    \end{equation}
    where $R_s$ being the shadow radius of the black as approximated in \cite{Hioki:2009na}:
\begin{equation}R_s=\frac{\left(X_b-X_r\right)^2+Y_t^2}{2|{X_r-X_t}|},
\end{equation}
which we obtain when we put $X_b=X_t$ and $Y_b=-Y_t$. The expression for the energy emission rate is given by
\begin{eqnarray}
\label{eq:energy_emission}
E_\omega=\frac{d^2E(\omega)}{d\omega dt}=\frac{2\pi^2 R_s^2}{e^{\omega/T_h}-1}\omega^3,
\end{eqnarray}
where $\omega$ being the photon frequency and $T_h$ is the Hawking temperature of the black hole event horizon. The expression of the temperature is written as--
\begin{eqnarray}
    \label{temp}
T_h=T_{\text{Kerr}}+\frac{\alpha  r_h-Q^2-Q_{\text{YM}}}{4 \pi  a^2 r_h+4 \pi  r_h^3}
\end{eqnarray}
where, $T_\text{Kerr}=\frac{r_h^2-a^2}{4 \pi  a^2 r_h+4 \pi  r_h^3}$ is the Kerr black hole temperature. For the rotating Yang-Mills inspired charged black hole surrounded by PFDM the expression of temperature is a Kerr-modified one, which means that we have the expression of the temperature for Kerr black hole and a correction term.
\begin{figure}[H]
    \centering
    \includegraphics[width=0.45\linewidth]{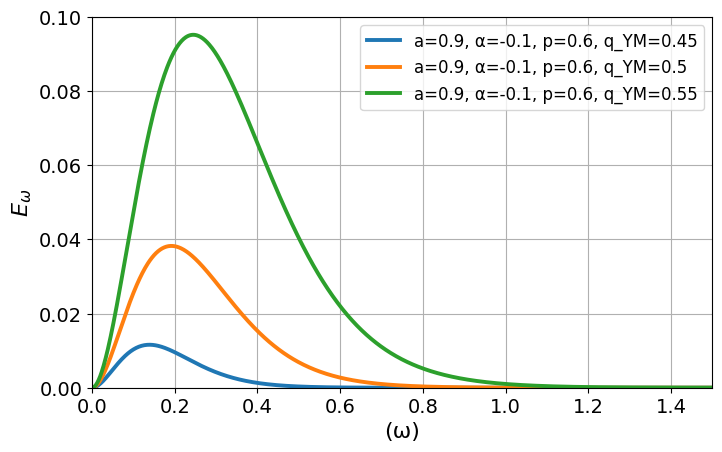}
    \includegraphics[width=0.45\linewidth]{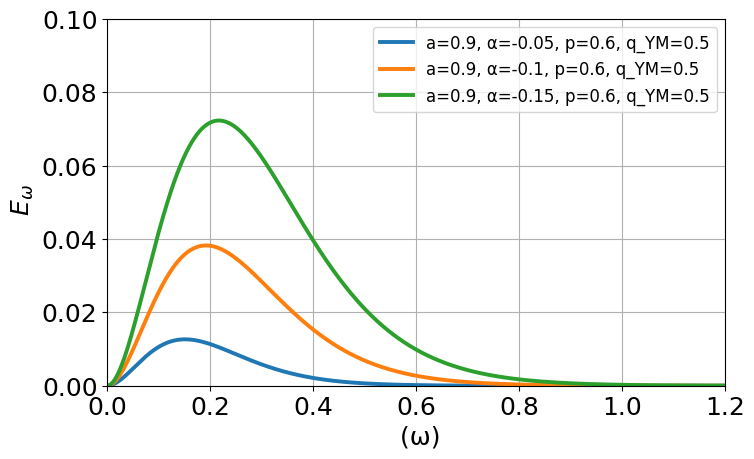}\\
    \caption{Evolution of the energy emission rate with the photon frequency $\omega$ for different set of values of $q_\text{YM}$ (left) and $\alpha$.}
    \label{energy_emission}
\end{figure}
In Fig.~\ref{energy_emission}, we depict the energy emission rate with respect to the photon frequency $\omega$, for different values of $q_{\text{YM}}$ for a fixed set of values of the spin and other parameters (left), and different values of $\alpha$ for a fixed values of $a$, $p$, and $q_{\text{YM}}$ (right). These plots suggest that with increasing values of $q_{\text{YM}}$, the energy emission rate increase and correspondingly, the peak of the Gaussian profile shifts to the higher values of $\omega$ (left). On the other hand, we have the similar structure of the energy emission profile with increasing values of $\alpha$.

\section{Summary and Discussion}\label{summary}
We study the possibility of existence of the Yang-Mills inspired charged black holes in the PFDM background by analyzing its shadow images and subsequent analyses of parameter estimations using the observables namely, the area ($A$) and oblateness ($D$). After the first ever detection of the images of shadows of M$87$ supermassive black holes by the EHT collaboration, there has been a surge of interest to the investigation of parametric bounds on various rotating black hole solutions coming from theories in general relativity and several modified theories of gravity. Although most of the astrophysical phenomena on the solar systems scale and beyond are consistence with the tests of general relativity, there are still certain possible tests which could be more robust and more accurate with the precision test astronomical observations. In this context, even though the EHT collaboration has modeled the Kerr black holes to analyze the shadow images of M$87$ and SgrA$^*$ supermassive black holes, the possible deviations from Kerr black holes and existence of non-Kerr geometries arising from modified gravity theories could not be avoided. \\
We speculate the numerical estimates of the observable parameters using the contour plots, namely, the shadow area ($A$), the oblateness ($D$) in the parameter spaces of $(a,q_\text{YM})$, $(a,\alpha)$, and $(a,p)$. We also determine the circularity deviation ($\Delta C$) of our concerned rotating spacetime with a modeling to M$87$ and SgrA$^*$ supermassive black holes and find the possible estimation of the pairs $(a,q_\text{YM})$, and $(a,\alpha)$. With the available methods as proposed in \cite{Kumar:2018ple}, we can estimate at least two parameters using either by $A$ or $\Delta C$ along with $D$. For example, we can determine $(a,q_\text{YM})$ or $(a,a)$ of the rotating Yang-Mills-modified charged black holes in PFDM environment. We model M$87$ supermassive black holes circularity deviation constraints, e.g., $\Delta C\leq0$ and found the bounds on the Yang-Mills charge $q_\text{YM}\leq 0.7$ and the PFDM parameter $\alpha\leq0.35$. Our analysis for M$87$ supermassive black holes is confined in the plane with the observer angles are at $\theta=17^\circ$ and $\theta=90^\circ$, respectively, while for SgrA$^*$ the observer are in the plane with angles  $\theta=50^\circ$ and $\theta=90^\circ$, respectively. Our analysis would be helpful for the analysis of shadow shapes and confining the parameters for various rotating black  hole systems in any gravity theories.

In effect, we also produced the results for the shadow deviation parameter $\delta$ and the shadow diameter $d_{sh}$ of our concerned rotating spacetime with modeling to M$87$ and SgrA$^*$ supermassive black holes. These analysis also put a more strict bounds on the parametric values characterizing black holes. Therefore, when the theoretically computed values of these observables with compared with those resulting from the astrophysical observations, we expect that one should know the complete and rigorous techniques used to determine information about our concerned black hole spacetimes. We also investigated the energy emission rate in connection with the black hole shadow radius, and the Hawking temperature. We plotted the energy emission rate as a function of photon frequency $\omega$, for a different set of values of $q_{\text{YM}}$, and $\alpha$.\\

A natural extension of this work would be to confront the shadow observables obtained for Yang--Mills--modified charged black holes in PFDM with the upcoming high-resolution data from next-generation EHT observations and space-based interferometry missions. Such comparisons will enable us to refine the bounds on $(a,q_{\text{YM}},\alpha)$ with unprecedented precision and to probe potential deviations from the Kerr paradigm.  \\
Another promising avenue is to extend the present analysis to dynamical scenarios, including time-dependent PFDM distributions and accretion environments, in order to assess their imprint on black hole shadow morphology and energy emission spectra. This would provide a more realistic framework to distinguish Yang--Mills--inspired black holes from their Kerr counterparts in observational astrophysics.

\section*{Acknowledgments}
AN acknowledges the support of Prof. Pankaj Joshi and the International Center for Space and Cosmology, Ahmedabad University, where part of this work was carried out during a summer internship. S.P would like to thank Prof. Hemwati Nandan for his support during the visit to HNB Garhwal University.

\section*{APPENDIX}
\appendix
\section{Rotating spacetime through the extended Newman-Janis algorithm}\label{appendix}
In this section, we discuss about the detailed calculations and algorithms to develop an axially symmetric rotating spacetime from a seed static spherically symmetric metric. The methods used to generate rotating  spacetimes were first developed in \cite{Newman:1965tw}. A detailed derivation of a more robust technique with or without complexifying the static spacetime with a similar rotating metric construction could be found from \cite{Demianski:1976tp}. We start with a static spherically symmetric metric of the form
\begin{eqnarray}
    \label{seed}
    ds^2=-f(r) dt^2+\frac{dr^2}{g(r)}+h(r)(d\theta^2+\sin^2\theta d\phi^2)
\end{eqnarray}
in which we partly apply the Newman-Janis algorithm. The first step in this direction would be to introduce the retarted null coordinates ($u,r,\theta,\phi$) so that
\begin{eqnarray}
    \label{retarted}
    du=dt-\frac{dr}{\sqrt{fg}}
\end{eqnarray}
The metric tensor is constructed from the null tetrads ($l^\mu,n^\mu,m^\mu,\bar{m}^\mu$) via,
\begin{eqnarray}
    \label{metricup}
    g^{\mu\nu}=l^\mu n^\nu+l^\nu n^\mu-m^\mu \bar{m}^\nu-m^\nu \bar{m}^\nu,
\end{eqnarray}
where
\begin{eqnarray}
    \label{null}
    &&l^\mu=\delta^\mu_r\\
     &&n^\mu=\sqrt{\frac{g}{f}}\delta^\mu_u-\frac{g}{2}\delta_r^\mu\\
     &&m^\mu=\frac{1}{\sqrt{2h}}\left(\delta_\theta^\mu+\frac{i}{\sin\theta}\delta_\phi^\mu\right)
\end{eqnarray}
and $$l^\mu l_\mu=m^\mu m_\mu=n^\mu n_\mu=l^\mu m_\mu=n^\mu m_\mu=0,\text{and}\,\,l^\mu n_\mu=-m^\mu\bar{m}_\mu=1$$

Next, we follow the complexification of the radial and null coordinates in the complex $r-u$ plane
\begin{eqnarray}
    \label{complexification}
    r\to r^\prime=r+ia\cos\theta,\;\text{and}\,u\to u^\prime=u-ia\cos\theta
\end{eqnarray}
Apply the above complexification to $\delta^\mu_\nu$ which transform like vectors as
\begin{eqnarray}
    \label{complexfication_1}
    \delta^\mu_r\to  \delta^\mu_r,\;  \delta^\mu_u\to \delta^\mu_u,\; \delta^\mu_\theta\to \delta^\mu_\theta+ia\sin\theta\left(\delta^\mu_u-\delta^\mu_r\right),\;\text{and}\,\,\delta^\mu_\phi\to \delta^\mu_\phi   
\end{eqnarray}
Consequently, we assume that the functions ($f,g,h$) also transform as ($F,G,H$) such that \cite{Azreg-Ainou:2014aqa,Azreg-Ainou:2014nra}
\begin{eqnarray}
    \label{complex functions}
    \lbrace f(r), g(r), h(r)\rbrace\to  \lbrace F(r,\theta,a),G(r,\theta,a), H(r,\theta,a)\rbrace
\end{eqnarray}
where $F(r,\theta,a),G(r,\theta,a),\;\text{and}\; H(r,\theta,a)$ are three real-valued functions. In our context we choose the functions in such a way that these satisfy the conditions \cite{Azreg-Ainou:2014aqa,Azreg-Ainou:2014nra,Azreg-Ainou:2014pra}
\begin{eqnarray}
    \label{limiting}
    \lim_{a\to 0} F(r,\theta,a)=f(r),\lim_{a\to 0} G(r,\theta,a)=g(r),\;\text{and}\; \lim_{a\to 0} H(r,\theta,a)=h(r)
\end{eqnarray}
These restrictions amount to the deviations from the usual Newman-Janis algorithm, which decides the functional forms of $F(r,\theta,a),G(r,\theta,a),\;\text{and}\; H(r,\theta,a)$ just by complexification of the radial coordinate $r$. For the procedure followed in Ref.~\cite{Azreg-Ainou:2014aqa,Azreg-Ainou:2014nra,Azreg-Ainou:2014pra}, the functional expressions, \{$F(r,\theta,a),G(r,\theta,a),H(r,\theta,a)$\} are fixed by using the criteria and other physical requirements.

The transformation procedure that affects the null tetrads  ($l^{\mu },n^{\mu},m^{\mu}$) as given in ~\eqref{complex functions} is also reflected on the conjugal  transformations of ~\eqref{complexfication_1} and~\eqref{complex functions} on $\delta_{\nu}^{\mu}$ and on \{$G(r),F(r),H(r)$\}, respectively:
\begin{align}
& l^{\mu } = \delta_r^{\mu}, \nonumber \\
\label{4n}& n^{\mu} = \sqrt{G/H} \,\delta^{\mu}_u - (G/2) \delta^{\mu}_r,  \\
& m^{\mu} = \big[ \delta_{\theta}^{\mu}+\text{i} a\sin\theta(\delta_u^{\mu}-\delta_r^{\mu}) + \frac{\text{i}}{\sin \theta}\,\delta_{\varphi}^{\mu} \big]/\sqrt{ 2 H}.
\end{align}
We can evaluate the inverse metric components as follows \cite{Azreg-Ainou:2014pra}--
\begin{align}
& g^{u u }(r,\theta) = -\frac{a^2 \sin^2\theta}{H}, \;\;
g^{u \varphi}(r,\theta) = -\frac{a}{H} , \nonumber  \\
& g^{\varphi \varphi}(r,\theta) = -\frac{1}{H\sin^2 \theta}  , \;\;
g^{\theta \theta}(r,\theta) = -\frac{1}{H},  \nonumber \\
& g^{rr}(r,\theta) = -G-\frac{a^2 \sin ^2 \theta }{H} , \;\;
g^{r \varphi}(r,\theta) = \frac{a}{H} ,\nonumber  \\
& g^{u r }(r,\theta) = \sqrt{\frac{G}{F}}+\frac{a^2 \sin ^2\theta }{H},
\end{align}
and hence the rotating metric in Eddington-Finkelstein Coordinates reads as \cite{Azreg-Ainou:2014aqa,Azreg-Ainou:2014nra,Azreg-Ainou:2014pra}
\begin{eqnarray}
\label{n5}
\rm d s^2 &&= F \rm du^2+2 \frac{\sqrt{A}}{\sqrt{B}} \rm d u \rm d r+2 a\sin ^2\theta  \Big(\frac{\sqrt{F}}{\sqrt{G}}-F\Big) \rm d u \rm d \varphi- 2 a\sin ^2\theta  \frac{\sqrt{F}}{\sqrt{G}} \rm d r \rm d \varphi\\&&-H  \rm d \theta^2
- \sin ^2\theta  \Big[H +a^2\sin ^2\theta  \Big(2 \frac{\sqrt{F}}{\sqrt{G}}-F\Big)\Big] \rm d \varphi^2.
\end{eqnarray}

Now as a last step we convert the metric ~\eqref{n5} to Boyer-Lindquist coordinates by a following global transformations of the coordinates which usually take the form
\begin{equation}
\label{n6}
\rm d u=\rm d t + \lambda(r)\rm d r,\,\rm d\varphi=\rm d\phi +\chi(r)\rm d r,
\end{equation}
we must ensure the integrability of the functions \{$\lambda,\chi$\} through their dependence on the radial coordinate $r$ only. We can easily integrate the above two equations to find the global coordinates $u(t,r)$ and $\varphi(\phi,r)$. Furthermore, the usual Newman-Janis Algorithm fails, in particular, to show ~\eqref{n5} to Boyer-Lindquist coordinates as in the usual Newman-Janis Algorithm, \{$F,G,H$\} are inherited just by complexification of the radial coordinate $r$. This way we do not have any free parameters or functions that act accordingly to achieve the transformation to Boyer-Lindquist coordinates. \\
In this procedure as the functions \{$F,G,H$\} are still not known and hence the transformation to Boyer-Lindquist coordinates can easily be achieved. This way we can get the functional form 
\begin{equation}\label{n7}
\lambda(r)=- \frac{(K+a^2)}{gh+a^2},\,\chi(r)=-\frac{a}{gh+a^2},
\end{equation}
where
\begin{equation}\label{n8}
K(r)\equiv \sqrt{\frac{g(r)}{f(r)}}h(r),
\end{equation}
the metric~\eqref{n5} is expressed in the usual Boyer-Lindquist coordinates provided we have
\begin{equation}\label{9}
F(r,\theta)=\frac{(gh+a^2\cos^2\theta) H}{(K+a^2 \cos^2\theta)^2},\, G(r,\theta)=\frac{gh+a^2\cos^2\theta}{H}.
\end{equation}
Hence, the desired form of the axisymmetric spacetime has the solution in Boyer-Lindquist coordinates as~\cite{Azreg-Ainou:2014aqa,Azreg-Ainou:2014nra,Azreg-Ainou:2014pra}
\begin{eqnarray}
\rm d s^2 &=& \frac{(gh+a^2 \cos ^2\theta )H \rm d t^2}{(K+a^2 \cos ^2\theta )^2}-\frac{H \rm d r^2}{gh+a^2}+2 a \sin ^2\theta\Big[\frac{K-gh}{(K+a^2 \cos ^2\theta )^2}\Big]H\rm d t\rm d \phi -H\rm d \theta^2\nonumber\\
&-&H\sin ^2\theta\Big[1+a^2\sin ^2\theta\frac{2K-gh
    +a^2\cos ^2\theta }{(K+a^2\cos ^2\theta )^2}\Big]\rm d \phi^2.
\label{n10}    
\end{eqnarray}
This latter metric is brought to Kerr-like forms~\cite{Azreg-Ainou:2014aqa,Azreg-Ainou:2014nra,Azreg-Ainou:2014pra}
\begin{eqnarray}
\rm d s^2 &=&\frac{\Psi}{\rho^2}\Big[\Big(1-\frac{2f}{\rho^2}\Big)\rm d t^2-\frac{\rho^2}{\Delta}\,\rm d r^2+\frac{4af \sin ^2\theta}{\rho^2}\,\rm d t\rm d \phi-\rho^2\rm d \theta^2-\frac{\Sigma\sin ^2\theta}{\rho^2}\,\rm d \phi^2\Big]
\label{10a}
\end{eqnarray}
\begin{eqnarray}
\label{10b}
\rm d s^2 &=&\frac{\Psi}{\rho^2}\Big[\frac{\Delta}{\rho^2}(\rm d t-a\sin^2\theta\rm d \phi)^2-\frac{\rho^2}{\Delta}\,\rm d r^2-\rho^2\rm d \theta^2-\frac{\sin^2\theta}{\rho^2}\,[a\rm d t-(K+a^2)\rm d \phi]^2\Big].
\end{eqnarray}
which are obtained when we perform the following variable changes:
\begin{align}
\label{10c}
&\rho^2\equiv K+a^2 \cos^2\theta,\;\;2f(r)\equiv K-gh\\
&\Delta(r)\equiv gh+a^2,\;\;\Sigma\equiv (K+a^2)^2-a^2\Delta\sin^2\theta.
\end{align}
To be specific, all the axisymmetric black holes in classical general relativity fulfill the requirements $g=f$ and $h=r^2$. With a particular choice of the functions in Eq.~\eqref{n8} implying that $K=H=r^2$, we can easily check that \cite{Azreg-Ainou:2014pra}
\begin{equation}\label{s2}
    \Psi=r^2+a^2\cos^2\theta
\end{equation}
Hence the rotating axisymmetric black hole spacetimes in the usual Boyer-Lindquist coordinates has the general form \cite{Azreg-Ainou:2014aqa,Azreg-Ainou:2014nra,Azreg-Ainou:2014pra}
\begin{eqnarray}
\label{r1}
\rm d s^2 &=&\Big(1-\frac{2f}{\rho^2}\Big)\rm d t^2-\frac{\rho^2}{\Delta}\,\rm d r^2+\frac{4af \sin ^2\theta}{\rho^2}\,\rm d t\rm d \phi-\rho^2\rm d \theta^2-\frac{\Sigma\sin ^2\theta}{\rho^2}\,\rm d \phi^2\\
\rho^2&=&r^2+a^2\cos^2\theta,\\
2f&=&r^2(1-F)\\
\Delta&=& r^2F+a^2=r^2-2f+a^2\\
\Sigma&=& (r^2+a^2)^2-a^2\Delta\sin^2\theta.
\end{eqnarray}
\bibliographystyle{JHEP}
\bibliography{pfdm_ym_black_hole}
\end{document}